\documentclass{elsart}
\usepackage{graphicx,amsmath,amssymb,mathptm}
\begin{document}
\begin{frontmatter}
\title{Evolution of scale-free random graphs: 
        Potts model formulation}
\author{D.-S.~Lee\thanksref{dslee}},
\thanks[dslee]{Present address: Theoretische Physik, Universit\"{a}t 
  des Saarlandes, 66041 Saarbr\"{u}cken, Germany}
\author{K.-I.~Goh, B.~Kahng, and D.~Kim}
\address{
School of Physics and Center for Theoretical Physics,
Seoul National University, Seoul 151-747, Korea}

\date{27 September 2004}
\begin{abstract}
We study the bond percolation problem in random graphs of $N$ weighted
vertices, where each vertex $i$ has a prescribed weight $P_i$ and an edge
can connect vertices $i$ and $j$ with rate $P_iP_j$.
The problem is solved by the $q\to 1$ limit of the $q$-state Potts 
model with inhomogeneous interactions for all pairs of spins. 
We apply this approach to the static model having 
$P_i\propto i^{-\mu} (0<\mu<1)$ so that the resulting graph is 
scale-free with the degree exponent $\lambda=1+1/\mu$. 
The number of loops as well as the giant cluster size 
and the mean cluster size are obtained in the thermodynamic 
limit as a function of the edge density, 
and their associated critical exponents are also obtained. 
Finite-size scaling behaviors are derived using
the largest cluster size in the critical regime, which is calculated from
the cluster size distribution, and checked against numerical simulation
results. 
We find that the process of forming the giant cluster is qualitatively
different between the cases of $\lambda >3$ and $2 < \lambda <3$. 
While for the former, the giant cluster forms abruptly at the 
percolation transition, for the latter, however, the formation of 
the giant cluster is gradual and the mean cluster size for 
finite $N$ shows double peaks.
 
\end{abstract}

\end{frontmatter}

\section{Introduction}
\label{sec:intro}
In the last few years graph theoretic approach has been of great 
value to characterize complex systems found in social, informational 
and biological areas. Here, a complex system is represented 
as a graph or network whose vertices and edges stand for its 
constituents and interactions. A simplest model for such is the random 
graph model proposed by Erd\H{o}s and R\'{e}nyi (ER)~\cite{er61}. 
In the ER model, $N$ number of vertices are present from the 
beginning and edges are added one by one in the system, 
connecting pairs of vertices selected randomly. Due to the randomness, 
the distribution of the number of edges incident on each vertex, 
called the degree distribution, is Poissonian. However, many 
real-world networks such as the World-wide web, the Internet, 
the coauthorship, the protein interaction networks and so on  
display power-law behaviors in the degree distribution. 
Such networks are called scale-free (SF) networks~\cite{ba99}. 
Thanks to recent extensive studies of SF networks, 
various properties of SF network structures have 
been uncovered~\cite{albert02,mendes02,newman03}. 

There have been a few attempts to describe scale-free networks in 
the framework of equilibrium statistical physics, even though the 
number of vertices grows with time in many real-world 
networks~\cite{berg02,manna03,burda,doro03,farkas}. 
In this approach, various mathematical tools developed in 
equilibrium statistical physics may be used to understand network 
structures. 
To proceed, one needs to define equilibrium network ensembles 
with appropriate weights, where one graph corresponds to one state 
of the ensemble. In a canonical ensemble, the number of edges $L$ is fixed: 
Given a degree distribution, $p_d(k)$, the mean degree 
$\langle k\rangle \equiv \sum k p_d(k)$ is obtained. 
Then the number of edges obtained through the relation, 
$L=\langle k\rangle N/2$, can be fixed. 
A degree sequence specifies the number of vertices with degree 
$k$ as $p_d(k)N$~\cite{molloy,newman01}.

A grandcanonical ensemble can be also defined, where 
the number of edges is also a fluctuating variable 
while keeping the SF nature of the degree distributions.   
The grandcanonical ensemble for SF random graphs is realized 
in the static model introduced by Goh {\it et al.}~\cite{goh01} or 
in its generalized version investigated in Refs.~\cite{caldarelli02}.
The name `static' originates from the fact that the number of 
vertices is fixed from the beginning. 
Here each vertex $i$ has a prescribed weight $P_i$ summed to 1 
and an edge can connect vertices $i$ and $j$ with rate 
$P_iP_j$~\cite{soderberg02,chung02,aiello02}.
A chemical potential-like parameter $K$ that can be regarded
as ``time" in the process of attaching edges controls the mean 
number of edges so that $\langle L\rangle$ increases with 
increasing $K$.
 
As the parameter $K$ increases, a giant cluster, or giant component, 
forms in the system. Here the giant cluster means the largest 
cluster of connected vertices whose size is ${\mathcal{O}}(N)$. 
Often such a giant cluster appears at the percolation 
transition point. In equilibrium statistical physics, 
the percolation problem can be studied through a spin model, 
the $q$-state Potts model in the $q\to 1$-limit~\cite{wu}. 
Using the relation, in this paper, we study the evolution of 
SF random graphs from the perspective of equilibrium statistical 
physics. To be specific, we construct the $q$-state Potts model, 
where the interaction strength between each pair of vertices is 
inhomogeneous on the complete graph. 
In this formulation, since the interaction strength $K$ is 
tunable, the mean number of edges $\langle L \rangle$ varies. 
Thus, the grandcanonical ensemble is taken in the network 
representation. However, since the number of spins (vertices) 
is fixed, the formulation corresponds to a canonical ensemble 
in the spin-model representation. Note that our model is different from 
the one studied by Dorogovtsev {\it et al.,}~\cite{potts_dorogo} where 
the Potts model is defined as a given fixed network so that 
each edge represents homogeneous interactions. 

The formulation of the spin model facilitates explicit derivation of various 
properties of the SF network. Thus we derive the formula for 
the giant cluster size, the mean cluster size, and in particular, 
the number of loops and clusters. These quantities are 
explicitly evaluated analytically for the static model with 
$P_i\propto i^{-\mu}$,
$(0<\mu<1)$ in the thermodynamic limit as a function of the edge 
density, and their critical properties are also studied. 
The degree exponent $\lambda$ is related to $\mu$ by 
$\lambda= 1+ 1/\mu $.
Moreover, their finite-size scaling behaviors 
are obtained using the {\it finite} largest cluster size for finite $N$   
that in turn is evaluated from the cluster size distribution.
From these, we are able to elucidate the process of formation of 
the giant cluster. While for the case $\lambda >3$, 
the giant cluster forms abruptly at the percolation transition point 
$K_c$, for the case $2<\lambda<3$ where most real world networks belong to, 
however, the formation of the giant cluster
is gradual and the mean cluster size for finite $N$ show double 
peaks.
 
In fact, the percolation problem of SF networks has been studied, 
but in a different way, that is, by removing vertices one by one 
as well as their attached edges from an existing SF 
network~\cite{cohen00,callaway00,cohen02}. 
The percolation transition was understood 
by using the branching process approach, which is supposed to be 
valid near the percolation transition point, where the network is 
sparse. 
In this paper, we 
provide the criterion for the 
validity of the branching process approach for a general degree 
distribution, and 
show that the branching process and the Potts 
model approaches are equivalent for the static model. 
Finally, note that while the branching process approach cannot 
count the number of loops, the Potts model formalism we use here 
enables us to count it. 

This paper is organized as follows. 
We introduce in Sec.~\ref{sec:static} an ensemble of random 
graphs where each vertex is weighted, 
and present in Sec.~\ref{sec:Potts}  
the Potts model formulation to derive graph theoretical 
quantities from its free energy.  
In Sec.~\ref{sec:bp},  the connection between the Potts model formulation
and the branching process approach
is discussed. The general results of Sec.~\ref{sec:Potts}
are applied to the 
static model in Sec.~\ref{sec:thermo} to obtain explicitly the 
giant cluster size, the mean cluster size and the mean number of loops
and clusters as a function of $K$. The cluster size distribution 
and the largest cluster size in finite size systems are obtained 
in Sec.~\ref{sec:distribution}.  The finite-size scaling is 
presented and compared with 
numerical simulation results in Sec.~\ref{sec:fss}. Finally  
Sec.~\ref{sec:summary} contains summary and discussion.

\section{Random graphs with weighted vertices}
\label{sec:static}

Suppose that the number of vertices $N$ is fixed (static) and each vertex 
$i=1,\ldots, N$ is given a probability $P_i$ summed to 1. 
The ER model of random graphs corresponds to assigning 
$P_i=1/N$ for all $i$. To construct a SF graph, we use 
$P_i \sim i^{-\mu}$ with $0<\mu<1$. 
However, for the time being, $P_i$ is arbitrary as long as $P_i\ll 1$ for 
all $i$.

In each unit time duration, two vertices $i$ and $j$ are selected 
with probabilities $P_i$ and $P_j$. If $i=j$ or an edge connecting 
$i$ and $j$ already exists, do nothing; otherwise, an 
edge is added between the vertices $i$ and $j$. This process is 
repeated for $NK$ times. 
Then the probability that a given pair of vertices 
$i$ and $j$ ($i\neq j$) is not connected by an edge is given by 
$(1-2P_i P_j)^{NK}\simeq e^{-2NKP_iP_j}$, while 
that it does is $1-e^{-2NKP_iP_j}$. Here we used the condition 
 $P_i\ll 1$. 
The factor $2$ comes from the equivalence of 
$(ij)$ and $(j i)$. We use the ``interaction'' parameter 
$K$ for later convenience which controls the edge density 
$\langle L \rangle/N$.  
Since each edge $b_{ij}$ is produced independently, 
this process generates a graph $G$ with probability 
\begin{eqnarray}
P(G)&=&\prod_{b_{ij}\in G} (1-e^{-2NKP_iP_j})\prod_{b_{ij}\notin G}e^{-2NKP_iP_j} 
\nonumber \\
&=&e^{-2NK\sum_{i>j}P_iP_j}\prod_{b_{ij}\in G} (e^{2NKP_iP_j}-1)\nonumber \\
&=&e^{-NK(1-M_2)}\prod_{b_{ij}\in G} (e^{2NKP_iP_j}-1),
\label{pg}
\end{eqnarray}
where we used the notation $M_n\equiv\sum_{i=1}^N P_i^n$.
By a graph $G$, we mean a configuration of undirected edges connecting 
a subset of $N(N-1)/2$ pairs of labelled vertices $i=1,2,\ldots,N$. 

We then evaluate the ensemble average of any graph theoretical quantity 
$A$ by 
\begin{equation}
\langle A \rangle =\sum_G P(G)A(G).
\end{equation}
One example is the degree $k_i$ of a vertex $i$, the number of edges  
incident on $i$. To do this, the generating function of $k_i$, 
$g_i(\omega)\equiv \langle \omega^{k_i}\rangle$, is first calculated 
as 
\begin{equation}
g_i(\omega)=\prod_{j(\ne i)} \left[e^{-2NKP_iP_j}+\omega 
(1-e^{-2NKP_iP_j})\right].
\label{giz}
\end{equation}
From this, one has 
\begin{equation}
\langle k_i \rangle = \left.\omega {d \over d\omega}g_i(\omega)\right|_{\omega=1}=
\sum_{j(\ne i)} (1-e^{-2NKP_iP_j}),
\label{kibn}
\end{equation}
and the average degree $\langle k\rangle$ is 
\begin{equation}
\langle k\rangle = {2\langle L\rangle\over N} =
{1\over N} \sum_{i} \langle k_i \rangle 
={1\over N}\sum_{i\ne j}(1-e^{-2NKP_iP_j}).
\label{lbn}
\end{equation}
Also, 
\begin{eqnarray}
\langle k_i^2\rangle&=&\left.
\left(\omega {d\over d\omega}\right)^2 g_i(\omega)\right|_{\omega=1}
\nonumber \\
&=&\left[\sum_{j(\ne i)} (1-e^{-2NKP_iP_j})\right]^2+
\sum_{j(\ne i)} e^{-2NKP_iP_j}(1-e^{-2NKP_iP_j})\nonumber \\
&=&\langle k_i \rangle +\langle k_i\rangle^2 -\sum_{j(\ne i)} (1-e^{-2NKP_iP_j})^2.
\label{ki2n}
\end{eqnarray}
We remark that Eq.~(\ref{giz}) is rewritten as 
\begin{equation}
g_i(\omega)\simeq e^{-(1-\omega) \langle k_i\rangle}
\label{gipoisson}
\end{equation}
with $\langle k_i \rangle$ in Eq.~(\ref{kibn}) 
when $|(1-\omega)(1-e^{-2NKP_iP_j})|\ll 1$ for all $j(\ne i)$. 
It implies that the probability that  $k_i$ is equal to $k$, 
$p_{d,i}(k)=\langle \delta_{k_i,k}\rangle$, is given by 
\begin{equation}
p_{d,i}(k)=\left.{1\over k!}{d^k \over d\omega^k} g_i(\omega)\right|_{\omega=0}
\simeq {\langle k_i\rangle^k \over k!} e^{-\langle k_i\rangle}
\end{equation}
for $k\gg 1$. 
Other quantities are discussed later on. 

\section{Potts model}
\label{sec:Potts}
\subsection{Potts model and random graph}
It is well known that the $q$-state Potts model provides a 
useful connection between the geometric bond percolation problem 
and the thermal systems through the Kasteleyn construction~\cite{wu}. 
The $q\to 1$ limit of the Potts model corresponds to the bond 
percolation problem. 
The same approach can be used for the random graph problem. From the 
viewpoint of the thermal spin system, this is basically the infinite 
range model since all pairs of spins interact with each other albeit with 
inhomogeneous interaction strength. 

Consider the $q$-state Potts Hamiltonian given by 
\begin{equation}
-H=2NK \sum_{i>j} P_i P_j \delta(\sigma_i,\sigma_j) + 
h_0 \sum_{i=1}^N [q\delta(\sigma_i,1)-1],
\label{hm}
\end{equation}
where $K$ is the interaction, $h_0$ is a symmetry-breaking field, 
$\delta(x,y)$  the Kronecker delta function, and 
$\sigma_i$ the Potts spins taking integer values $1,2,\ldots, q\equiv r+1$. 
We use the notation $r\equiv q-1$. 
The partition function $Z_N(q,h_0)$ can be written as 
\begin{eqnarray}
Z_N(q,h_0)&=&{\rm Tr}\, e^{-H} = 
{\rm Tr} \, \prod_{i>j} \left[
1+(e^{2NKP_iP_j}-1) \delta(\sigma_i,\sigma_j)\right] \nonumber \\
&&\times        \prod_i e^{h_0 (q\delta(\sigma_i,1)-1)},
\end{eqnarray}
where Tr denotes the sum over $q^N$ spin states. Expanding the first 
product and taking the Tr operation, one has 
\begin{equation}
Z_N(q,h_0)=\sum_G \prod_{b_{ij}\in G} (e^{2NKP_iP_j}-1) 
\prod_{s\geq 1} (e^{srh_0}+r e^{-sh_0})^{n_G(s)},
\label{zqh1}
\end{equation}
where $n_G(s)$ is the number of $s$-clusters, a cluster with $s$ vertices 
in a given graph $G$. Comparing this with Eq.~(\ref{pg}), one immediately notices that 
\begin{equation}
Z_N(q,h_0)=e^{NK(1-M_2)} \sum_G P(G) 
\prod_{s\geq 1} (e^{srh_0}+re^{-sh_0})^{n_G(s)}.
\end{equation}
In particular, $Z_N(q,0)=e^{NK(1-M_2)} \langle q^C\rangle$, 
where $C=\sum_s n_G(s)$ is the total number of clusters in graph $G$.  
Thus $Z_N(q,0)$ is the generating function of $C$. 

The magnetization of the Potts model is defined as 
\begin{equation}
m(q,h_0)
={1\over rN} {\partial \over \partial h_0} \ln Z_N(q,h_0). 
\end{equation}
It can be written as 
\begin{equation}
m(q,h_0)=
{\left\langle \sum_{\rm clusters} \left[
        {e^{srh_0}-e^{-sh_0} \over e^{srh_0}+r e^{-sh_0}}
        \right] 
\left({s\over N}\right)
\prod_{\rm clusters}(
        e^{srh_0}+r e^{-sh_0}) \right\rangle \over 
\langle \prod_{\rm clusters} (
        e^{srh_0}+r e^{-sh_0}) \rangle
}.
\end{equation}

If we introduce the cluster size distribution 
$P(s)\equiv n(s)(s/N)$ with $n(s)=\langle n_G(s)\rangle$ 
and the generating function $\mathcal{P}(z)=\sum_{s\ge 1}
P(s) z^s$,  
the magnetization is, when $q=1$,  
\begin{equation}
m(1,h_0)=\sum_{s\ge 1} P(s) 
\left(1-e^{-sh_0}\right) =
1-\mathcal{P}(e^{-h_0}).
\end{equation}
The generating function $\mathcal{P}(z)$ will be used in
Sec.~\ref{sec:distribution} 
to investigate the  asymptotic 
behavior of the cluster size distribution.  

When $h_0=0$, the magnetization  vanishes for finite $N$. 
However, 
when we take the limit $h_0\to 0$ after the thermodynamic limit $N\to\infty$, 
the contribution from the largest cluster whose size is $S$    
can survive to give  
\begin{equation}
m(1,h_0\to 0)=\left\langle {S\over N}\right\rangle,
\label{m10}
\end{equation}
if $S/N$ is finite.
Let us define  a giant cluster by a cluster whose size is $\mathcal{O}(N)$. 
Then $m(1,h_0\to 0)$ is the ratio of the giant cluster size to $N$, if 
it exists, and the system is considered as being in the percolating 
phase if $m(1,h_0\to 0)$ is non-zero.
For simplicity, we will call $m(1,h_0\to 0)$ the giant cluster size 
and denote it by $m$. 

The susceptibility defined as 
$\chi(q,h_0)\equiv (1/q)(\partial/\partial h_0) m(q,h_0)$ 
on the other hand is  related to the mean cluster size:
\begin{equation}
\chi(1,h_0\to 0)=\lim_{h_0\to 0} \lim_{N\to \infty}\sum_s P(s) s e^{-sh_0} = 
\sum_{s\ne \langle S\rangle} sP(s), 
\label{chi}
\end{equation}
where $h_0 \langle S\rangle \to \infty$ is used with $\langle S\rangle$ 
the ensemble average of the largest cluster size, which we call simply 
the largest cluster size.
We will denote $\chi(1,h_0\to 0)$ by $\bar{s}$. 
Note that our definition of $\bar{s}$ is normalized with 
respect to the total number of vertices instead of the number of 
vertices belonging to finite clusters. 

The number of loops $N_{\rm loop}$ is related to the 
total number of  clusters $C$ through
\begin{equation}
N_{\rm loop}=L-N+C.
\label{lnc}
\end{equation}
Since $Z_N(q,0)\sim \langle q^C \rangle$, one can notice that 
the number of loops per vertex $\langle N_{\rm loop}\rangle/N$
is given as 
\begin{equation}
{\langle N_{\rm loop}\rangle \over N}={\langle L\rangle\over N}-1
+{1\over N}{\partial \over \partial q}\left[
\ln Z_N(q,0)\right]_{q=1}.
\end{equation}
We will denote $\langle N_{\rm loop}\rangle/N$ by $\ell$ and 
call it the number of loops for simplicity.

\begin{figure}[t]
\centering
\includegraphics[width=8.5cm]{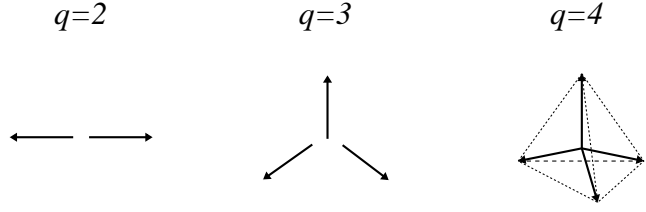}
\caption{Vector representations of $q$-state Potts spins with 
        $q=2$, $3$, and $4$.}
\label{fig:vector}
\end{figure}

\subsection{Partition function}
A convenient way to evaluate the partition function 
is to resort to the vector-spin representation where 
one associates an $r$-dimensional vector $\vec{S}(\sigma_i)$ of unit 
length to each spin value $\sigma_i$, 
where $\vec{S}(1)=(1,0,\ldots,0)$ and $\vec{S}(\sigma_i)$ with $\sigma_i=2,3,
\ldots, q$ point to the remaining $r$ corners of the $r$-dimensional 
tetrahedron (See Fig.~\ref{fig:vector}).
Then one can represent the Kronecker delta function as a dot product 
between $\vec{S}$'s,
\begin{equation}
\delta(\sigma_i,\sigma_j)={1\over q}(r\, \vec{S}(\sigma_i)\cdot 
\vec{S}(\sigma_j)+1).
\end{equation}
Using this, the interaction term in Eq.~(\ref{hm}) can be written as 
\begin{equation}
2NK\sum_{i>j}P_i P_j \delta(\sigma_i,\sigma_j)=
NK\left({1\over q}-M_2\right)+{rNK\over q}  
\left(\sum_i P_i \vec{S}(\sigma_i)\right)^2.
\end{equation}
The perfect square is then linearized through the identity 
$\int dy e^{-ay^2+by}=\sqrt{\pi/a}\, e^{b^2/(4a)}$. Thus we have 
\begin{equation}
Z_N(q,h_0)={\rm Tr}\, \left[e^{NK(\frac{1}{q}-M_2)} 
\left({4\pi K\over rqN}\right)^{-\frac{r}{2}} 
e^{\sum_{i=1}^N r\vec{h_0} \cdot \vec{S}(\sigma_i)}
\int d\vec{y}\,  e^{-{rqN\over 4K} y^2 + rN\sum_{i=1}^N P_i \vec{y}\cdot 
\vec{S}(\sigma_i)}  \right], 
\end{equation}
where the integration is over the $r$-dimensional space and 
$\vec{h_0}=(h_0,0,\ldots,0)$.
Now the Tr operation can be performed for each spin independently.
Defining 
\begin{equation}
\tilde{\zeta}(\vec{h})={1\over q}
\sum_{\sigma=1}^q e^{r\vec{h} \cdot \vec{S}(\sigma)},
\end{equation}
one then has 
\begin{equation}
Z_N(q,h_0)=
\left({4\pi K\over rqN}\right)^{-\frac{r}{2}}  q^N e^{NK(\frac{1}{q}-M_2)}
\int d\vec{y} \, e^{-{rqN\over 4K} y^2 + \sum_{i=1}^N \ln 
\tilde{\zeta}(\vec{h_0}+NP_i\vec{y})}.
\label{zqh2}
\end{equation}

Provided $(1/N)\sum_{i=1}^N \ln \tilde{\zeta}(\vec{h_0} +NP_i\vec{y})$ 
has a well-defined limit as $N\to\infty$, one can apply the saddle point 
method to Eq.~(\ref{zqh2}), where the integral is replaced by the value 
of the integrand at its maximum. The maximum is for $\vec{y}$ which 
is a solution of the saddle-point equation
\begin{equation}
{q\over 2K} \vec{y}={1\over rN} \sum_{i=1}^N \nabla_{\vec{y}} 
\ln \tilde{\zeta}(\vec{h_0}+NP_i\vec{y}).
\label{sp1}
\end{equation}
When $\vec{h}_0=0$, $\vec{y}=0$ is always a solution of Eq.~(\ref{sp1}), 
but the spontaneous symmetry-breaking 
solutions ($\vec{y} \ne 0$) with the Potts symmetry may appear for large $K$. 
When the symmetry-breaking field is applied along the $1$-direction, 
$\vec{h}_0=(h_0,0, \ldots,0)$, 
the nontrivial physically relevant solution of Eq.~(\ref{sp1}) is expected 
in the sub-manifold $\vec{y}=(y,0,\ldots,0)$.  The limit $h_0 \rightarrow 0^+$
then selects one of the $q$ equivalent spontaneous symmetry-breaking solutions. 
With this in mind, we may restrict our attention to the one-dimensional sub-manifold of  
$\vec{y}$ in Eqs.~(\ref{zqh2}) and (\ref{sp1}).  
As a result, we then have, as $N\to\infty$,
\begin{equation}
{1\over N}\ln Z_N(q,h_0)=\ln q +K \left({1\over q}-M_2\right)-r F(y,h_0),
\label{logZ}
\end{equation}
with
\begin{equation}
F(y,h_0)={q\over 4K}y^2 - {1\over rN}\sum_{i=1}^N \ln \zeta(h_0+NP_iy,q),
\label{fyh0}
\end{equation}
where
\begin{equation}
\zeta(h,q)={1\over q}\sum_{\sigma=1}^q e^{r S_1(\sigma)} = 
{e^{rh}+r e^{-h}\over 1+r},
\label{zetah}
\end{equation}
and $y$ is the solution of the one-dimensional saddle-point equation, 
\begin{equation}
{q\over 2K} y ={1\over r}\sum_{i=1}^N P_i {\partial\over \partial h_0} 
\ln \zeta(h_0+NP_i y,q).
\label{sp2}
\end{equation}
Here, the $q$-dependence in $F(y,h_0)$ is not shown explicitly. 
Since $m=(1/(rN)) (\partial/\partial h_0) \ln Z$, we see that 
\begin{equation}
m(q,h_0)=-{d \over dh_0}F(y,h_0)=
{1\over rN}\sum_{i=1}^N {\partial \over \partial h_0} 
\ln \zeta(h_0+NP_iy,q).
\label{mag}
\end{equation}
At this point, it is useful to take the $r\to 0$ limit in 
Eqs.~(\ref{fyh0}), (\ref{sp2}), and (\ref{mag}), which yields, with $h_i=h_0+NP_iy$, 
\begin{equation}
F(y,h_0)={1\over 4K}y^2-{1\over N}\sum_{i=1}^N (e^{-h_i}-1+h_i), 
\label{fyh02}
\end{equation}
where $y$ is the solution of 
\begin{equation}
{y\over 2K}=\sum_{i=1}^N P_i (1-e^{-h_i}). 
\label{sp3}
\end{equation}
The magnetization and the susceptibility reduce to 
\begin{equation}
m(1,h_0)={1\over N}\sum_{i=1}^N (1-e^{-h_i}),
\label{mag2}
\end{equation}
and
\begin{eqnarray}
\chi(1,h_0)&=&{1\over N}\sum_{i=1}^N e^{-h_i} 
(1+NP_i {dy \over dh_0}), \nonumber \\
&=&
{1\over N}\sum_{i=1}^N e^{-h_i} + 
{\left(\sum_{i=1}^N P_i e^{-h_i}\right)^2 \over 
        (2K)^{-1}-\sum_{i=1}^N NP_i^2 e^{-h_i}},
\label{chi2}
\end{eqnarray}
respectively, where it is used that   
\begin{equation}
{dy\over dh_0}={\sum_{i=1}^N P_i e^{-h_i} \over 
        (2K)^{-1}-\sum_{i=1}^N NP_i^2 e^{-h_i}}.
\end{equation}
Thus the giant cluster size and the mean cluster size are obtained
from Eqs.~(\ref{mag2}) and (\ref{chi2}), respectively, with $h_0\to 0$.

Also, the number of clusters per vertex is 
\begin{equation}
{\langle C \rangle\over N}= 1-K
-F(y,0)|_{q=1},
\label{covern}
\end{equation}
while that of 
loops is  
\begin{equation}
{\langle N_{\rm loop}\rangle\over N}= {\langle L \rangle \over N }
-K -F(y,0)|_{q=1},
\label{ell1}
\end{equation}
where $y$ is given by Eq.~(\ref{sp3}) with $h_0=0$.

When $h_0\to 0$, a non-trivial solution of Eq.~(\ref{sp3}) 
begins to appear when $(2K)^{-1}<N\sum_{i=1}^N P_i^2$,  which 
gives the following characteristic value $K_c$ 
\begin{equation}
K_c={1 \over 2N\sum_{i=1}^N P_i^2}. 
\label{kc0}
\end{equation}
When $P_i$ decays slower than $i^{-1}$ and $KP_i\ll 1$ for all $i$, 
$\langle k\rangle=2K$ and 
$\langle k^2 \rangle=(1/N)\sum_{i=1}^N \langle k_i^2\rangle= \langle k\rangle
+N^{-1}\sum_{i=1}^N \langle k_i\rangle^2=
2K+4NK^2 \sum_{i=1}^N P_i^2$, which will be shown below. 
Then the condition $K=K_c$ is equivalent to 
the well-known condition 
$\langle k^2 \rangle/\langle k\rangle=2$~\cite{molloy}. 
Whether the percolation transition occurs at $K_c$ or not 
will be investigated for specific $P_i$'s of the static model. 

\section{Branching process approach}
\label{sec:bp}
The cluster size distribution $P(s)$ 
can be obtained from Eqs.~(\ref{sp3})  and (\ref{mag2}) using  
$m(1,h_0)=1-\mathcal{P}(e^{-h_0})$ and $P(s)=(1/s!)d^s\mathcal{P}(z)/dz^s|_{z=0}$. 
However, 
the cluster size distribution can also be obtained 
through the generating function approach or equivalently the 
branching process approach. 
Here, the presence of loops in {\it finite} clusters is neglected and  
each cluster in a given graph is considered as a tree 
generated by successive branchings from an arbitrary vertex
~\cite{harris89,aharony92}. 
Consider the probability that a randomly chosen vertex belongs to 
a $s$-cluster, which is just $P(s)$. 
Then $P(s)$ can be written recursively as 
\begin{equation}
P(s)=\delta_{s,1}p_d(0)+
\sum_{k\ge 1}p_d(k) \prod_{i=1}^k\sum_{s_i}R(s_i)\,\delta_{\sum_i s_i,s-1}, 
\label{prec}
\end{equation}
where $p_d(k)$ is the degree distribution and $R(s)$ 
is the probability that a randomly-chosen edge has   
a $s$-cluster at its one end, and thus equal to the 
 number of edges followed by $s$-clusters divided by $2L$. 
$R(s)$ is obtained self-consistently as 
\begin{equation}
R(s)=\delta_{s,1}r_d(0)+
\sum_{k\ge 1}r_d(k) \prod_{i=1}^k\sum_{s_i}R(s_i)\,\delta_{\sum_i s_i,s-1}, 
\label{rrec}
\end{equation}
where $r_d(k)$ is the probability that the vertex at either end of 
a randomly-chosen edge has $k+1$ edges and thus is equal to 
$(k+1)p_d(k+1)/\langle k\rangle$. 
With the generating functions $\mathcal{P}(z)=\sum_{s=1}^\infty 
P(s) z^s$ and $\mathcal{R}(z)=\sum_{s=1}^\infty R(s) z^s$, 
Eqs.~(\ref{prec}) and (\ref{rrec}) can be written in more compact forms as 
\begin{equation}
\mathcal{P}(z)=z g(\mathcal{R}(z))
\label{psc}
\end{equation}  
and
\begin{equation}
\mathcal{R}(z)=z f(\mathcal{R}(z)),
\label{rsc}
\end{equation}
where $g(\omega)=\sum_{k=0}^\infty p_d(k)\omega^k$ and 
$f(\omega)=g'(\omega)/\langle k \rangle = g'(\omega)/g'(1)$.
We mention that Eqs.~(\ref{psc}) and (\ref{rsc}) 
with $z=1$ are equivalent to those derived by Molloy and Reed~\cite{molloy}
for a given degree sequence. 

Eqs.~(\ref{psc}) and (\ref{rsc}) are the standard results. 
For the grandcanonical ensemble we are using, 
the generating functions $g(\omega)$ and $f(\omega)$ are represented 
in terms of $g_i(\omega)$, the generating function of $p_{d,i}(k)$ 
in Eq.~(\ref{giz}),  as 
\begin{eqnarray}
g(\omega)&=&{1\over N}\sum_{i=1}^N g_i(\omega) 
        \nonumber \\
f(\omega)&=&{1\over N\langle k \rangle}\sum_{i=1}^N {dg_i (\omega)\over
        d\omega}
\label{fg}
\end{eqnarray}
with $\langle k \rangle=g'(1)$. 
In particular, if 
\begin{equation}
g_i(\omega)= e^{-(1-\omega)2NKP_i}, 
\label{bpvalid}
\end{equation}
which holds, for example, when $1-e^{-2NKP_iP_j}\ll 1$ for all $i\ne j$, 
then Eqs.~(\ref{psc}) and (\ref{rsc}) 
of the branching process approach 
are exactly equal to Eqs.~(\ref{mag2}) and (\ref{sp3}) of 
the Potts model formulation,  
identifying $z$, $\mathcal{P}(z)$, and  
$\mathcal{R}(z)$ with $e^{-h_0}$, $1-m(1,h_0)$, 
and $1-y/(2K)$, respectively. 

\section{Percolation of the Static model: Thermodynamic limit}
\label{sec:thermo}

\begin{figure*}
\centering
\includegraphics[width=14cm]{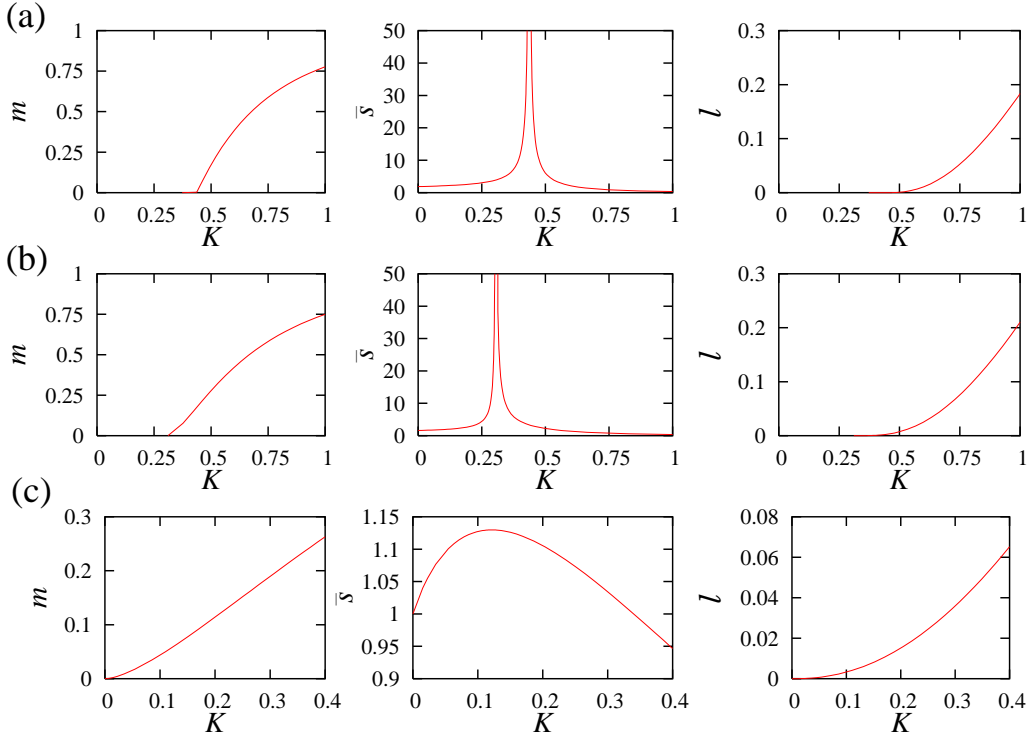}
\caption{Giant cluster size $m$, mean cluster 
        size $\bar{s}$, and number of loops $\ell=\langle 
        N_{\rm loop}\rangle/N$ vs. $K$ for $\mu=5/19$ 
   ($\lambda=4.8$) (a), $\mu=5/13$ ($\lambda=3.6$) (b), and $\mu=5/7$
   ($\lambda=2.4$) (c). 
        They are obtained by solving numerically Eqs.~(\ref{mag3}), (\ref{chi3}), 
        and (\ref{ell2}), respectively, together with Eq.~(\ref{ysc}). 
The number of clusters per vertex not shown here is $1-K+\ell$ and 
monotonically decreases with $K$ from 1 to 0.} 
\label{fig:num_sol}
\end{figure*}

So far, our discussion applies to arbitrary $P_i$. In this section, 
we specialize to  the case of the static model, 
\begin{equation}
P_i={i^{-\mu}\over \zeta_N(\mu)} \quad (0<\mu<1).
\label{Pi}
\end{equation}
Here
$\zeta_N(x)\equiv \sum_{i=1}^N i^{-x}$, and in the limit $N\to\infty$,
it converges to the Riemann zeta function $\zeta(x)$ when $x>1$ and diverges as
$N^{1-x}/(1-x)$  when $0<x<1$.
In the marginal case with $x=1$, it is calculated as
$\ln N + \gamma_M$ with $\gamma_M=0.5772\ldots$, the Euler-Mascheroni constant.
The sum appearing in Eq.~(\ref{fyh02}) is evaluated in Appendix A as
\begin{eqnarray}
\Sigma_1(y)&\equiv&{1\over N}\sum_{i=1}^N e^{-NP_i y}\nonumber \\
&=& - \Gamma\left(1-{1\over
                \mu}\right)(1-\mu)^{1\over \mu} y^{1\over \mu}+1-y
+{(1-\mu)^2 \over 2(1-2\mu)} y^2 + \cdots 
\label{sigma1c}
\end{eqnarray}
for $y(1-\mu)N^\mu\gg 1$. 
This will be used repeatedly. We do not consider the marginal cases
where $1/\mu$ is an integer. 
The sums in Eqs.~(\ref{kibn}) and (\ref{lbn}) 
are evaluated using $\Sigma_1(y)$ to give 
\begin{equation}
\langle k_i \rangle = 2NK P_i \quad {\rm and} \quad 
\langle k \rangle = 2\langle L\rangle/N=2K,
\label{kik}
\end{equation}
in the limit $N\to\infty$ if $KP_i\ll 1$ for all $i$ or equivalently, 
$K\ll \zeta_N(\mu)=\mathcal{O}(N^{1-\mu})$. 
Thus $K$ is $\langle L\rangle /N$. 
Under the same condition, the third term  
$(1-e^{-2NKP_iP_j})^2$ in Eq.~(\ref{ki2n}) 
does not contribute to $\langle k^2 \rangle=(1/N)\sum_{i=1}^N\langle k_i^2\rangle$ 
in the limit $N\to\infty$, which gives $\langle k^2\rangle=(1/N)\sum_{i=1}^N
[\langle k_i\rangle+\langle k_i\rangle^2]$. 
Moreover, rewriting Eq.~(\ref{giz}) as 
\begin{equation}
\log g_i (\omega) = \sum_{j (\neq i)} \log \left[
       1-(1-\omega) (1- e^{-2NK P_i P_j}) \right]
\label{giz2} 
\end{equation}
and expanding the right hand side as a power series in 
$(1-\omega)$ to apply the result of Appendix A, we find 
that Eq.~(\ref{bpvalid}) holds
for all range of $0 < \mu< 1$ and $K$ finite. Note that 
$(1-e^{-2NKP_i P_j})$ is not small when $1/2<\mu<1$ but the final 
result is the same as that for $0<\mu<1/2$ where $(1-e^{-2NKP_i P_j})
\simeq 2NKP_i P_j \ll 1$ holds. Thus the degree of each vertex $k_i$
follows the Poisson distribution and the branching process
approach and the Potts model approach are equivalent for the 
static model as long as $m$ and ${\bar s}$ are concerned.  
This is because the {\it finite} clusters remain effectively  
trees for all $K$.

For convenience, we divide the range of $\mu$ into 
the three cases, (I) $0<\mu<1/3$, (II) $1/3<\mu<1/2$, 
and (III) $1/2<\mu<1$.

\subsection{Degree distribution} 

The asymptotic behavior of the degree distribution $p_d(k)$ is  
related to the behavior of its generating function $g(\omega)=\sum_k
p_d(k)\omega^k$ for $\omega$ near $1$, which is 
equal to $\Sigma_1(2K(1-\omega))$ in the limit $N\to\infty$. 
The latter condition is necessary for the approximation 
$P_i(1-P_i)\simeq P_i$ to be valid. 
From  Eqs.~(\ref{sigma1o}) and
(\ref{sigma1b}),  the degree distribution $p_d(k)$ 
is given by 
\begin{equation}
p_d(k)={1\over k!}\left.{d^k\over d\omega^k} g(\omega)\right|_{\omega=0}\simeq\left\{
\begin{array}{ll}
c_1 k^{-1-\frac{1}{\mu}}&{\rm for}\ 1\ll k\ll k_{\rm max},\\[2.5mm]
c_2 \frac{k_{\rm max}^k}{k!}&{\rm for} \ k\gg k_{\rm max},
\end{array}
\right.
\label{degree}
\end{equation}
where $k_{\rm max}$ is equal to $\langle k_1 \rangle$ i.e.,  
$k_{\rm max}=2K(1-\mu)N^{\mu}$,  
$c_1=(1/\mu)[2K(1-\mu)]^{1/\mu}$ and $c_2
        =(1/N)\sum_{r=0}^\infty (-k_{\rm max})^r \zeta[\mu(k+r)]/r!\simeq e^{-k_{\rm max}}/N$. 
From now on, we assume that $K_{\ell}\ll K\ll K_u$ 
with $K_{\ell}\equiv N^{-\mu}/(2(1-\mu))$ 
and $K_u\equiv N^{1-\mu}/(2(1-\mu))$, for which 
$1\ll 
k_{\rm max}\ll N$ so that 
there exists the regime of $k$ where the degree distribution follows 
a power law 
$p_d(k)\sim k^{-\lambda}$ with 
\begin{equation}
\lambda=1+{1\over \mu}.
\end{equation} 
Since we are interested in the range $0< \mu<1$, 
the degree exponent $\lambda$ is larger than $2$.

\subsection{Giant cluster size}
The giant cluster size $m$ 
can be evaluated by Eq.~(\ref{mag2}) with $h_i=NP_iy$. 
In terms of $\Sigma_1(y)$ evaluated in 
Appendix A, 
it is simply represented  as 
\begin{equation}
m=1-\Sigma_1(y)
\label{mag3}
\end{equation}
with $y$ obtained by solving Eq.~(\ref{sp3})   
\begin{equation}
{y\over 2K}=\sum_{i=1}^N P_i (1-e^{-h_i})
=1+\Sigma_1'(y),
\label{ysc}
\end{equation}
where $\Sigma_1'(y)=(d/dy)\Sigma_1(y)$. 

When $y$ is small, Eq.~(\ref{ysc}) is expanded as 
\begin{eqnarray}
({\rm I})~~~
{y\over 2K}&\simeq& {y\over 2K_c}-{(1-\mu)^3\over 2 (1-3\mu)} y^2,
        \nonumber\\[2.5mm]
({\rm II})~~~
{y\over 2K}&\simeq& {y\over 2K_c}-\Gamma\left(1-\frac{1}{\mu}\right)
{(1-\mu)^{1\over \mu} \over \mu} y^{\frac{1}{\mu}-1}, \nonumber \\[2.5mm]
({\rm III})~~~
{y\over 2K}&\simeq& -\Gamma\left(1-\frac{1}{\mu}\right)
{(1-\mu)^{1\over \mu} \over \mu} y^{\frac{1}{\mu}-1},
\label{ysc_expand}
\end{eqnarray}
for the three ranges of $\mu$,  (I), (II), and (III), respectively.  
For (I) and (II), the  characteristic value $K_c$ defined by  
\begin{equation}
K_{c}={(1-2\mu)\over 2 (1-\mu)^2}
\label{kc}
\end{equation}
appears, which is just Eq.~(\ref{kc0}) with $P_i$ in Eq.~(\ref{Pi}).

When $K<K_c$ for (I) and (II), or $K=0$ for (III),  
Eqs.~(\ref{ysc}) or (\ref{ysc_expand}) has the solution $y=0$, and therefore, 
the giant cluster size is 
\begin{equation}
m=0.
\end{equation}
That is, there is no giant cluster for $K<K_c$ (I, II) or $K=0$ (III).
But a non-zero solution for $y$  occurs 
when $K>K_c$ (I, II) or $K>0$ (III).
It leads to the following giant cluster size $m$:
\begin{eqnarray}
({\rm I})~~~&m\simeq& y\simeq {2(1-3\mu)\over
        (1-\mu)(1-2\mu)}\Delta, \nonumber \\[2.5mm]
({\rm II})~~~&m\simeq&y\simeq {1\over 1-\mu}\left(
         {\mu \over (1-2\mu)\Gamma\left(1-\frac{1}{\mu}\right)}
                \right)^{\mu\over 1-2\mu} \Delta^{\mu\over 1-2\mu},\nonumber \\
({\rm III})~~~&m\simeq&y\simeq \left({2(1-\mu)^{1\over \mu}
	  \left|\Gamma\left(1-\frac{1}{\mu}\right)\right|\over \mu}\right)^{\mu\over 2\mu-1} K^{\mu\over 2\mu-1},
\label{ysol}
\end{eqnarray}
for  $\Delta\equiv K/K_c-1$ (I, II) or $K$ (III) small and positive. 
Here the relation $m\simeq y$ comes from Eqs.~(\ref{mag3}) and (\ref{sigma1b}). 
The giant cluster size is finite for $\Delta$ finite and positive, 
and thus $K_c$ is the percolation transition point. 
If we define a critical exponent $\beta$ by $m\sim \Delta^\beta$, its value is $1$ (I) and  $\mu/(1-2\mu)$
                (II). 
For (III), 
$m$ is finite for $K$ finite, but $K=0$ is not a percolation transition 
point, which will be investigated further below. 

The giant cluster size $m$ as a function of $K$, 
which can be obtained numerically from Eqs.~(\ref{mag3}) and (\ref{ysc}), 
is plotted for the case of $\mu=5/19$ ($\lambda=1+1/\mu=4.8$), 
$5/13$ ($\lambda=3.6$), and $5/7$ ($\lambda=2.4$)  
in Fig.~\ref{fig:num_sol}.  

We mention that in Ref.~\cite{chung02}, 
some rigorous bounds of the giant cluster size are derived 
for an ensemble similar to ours, but with a different form of 
the probabilities of adding edges so that their results 
apply only to the case $\lambda >3$ of the static model. 

\subsection{Mean cluster size}
The mean cluster size $\bar{s}$ or the susceptibility $\chi(1,h_0\to 0)$  
in Eq.~(\ref{chi2}) 
is represented in terms of $\Sigma_1(y)$ in Appendix A as 
\begin{equation}
\bar{s}=\Sigma_1(y)+{[\Sigma_1'(y)]^2 
        \over (2K)^{-1}-\Sigma_1''(y)},
\label{chi3}
\end{equation}
where $y$ is the solution of Eq.~(\ref{ysc}) and 
$\Sigma_1''(y)=(d/dy)\Sigma_1'(y)$. 

When $K<K_c$ for (I) and (II), 
$y=0$ and thus 
\begin{equation}
\bar{s}=1+{2KK_c\over K_c-K}
\end{equation}
since $\Sigma(0)=-\Sigma'(0)=1$, and $\Sigma_1''(y) = 1/(2K_c)$.
On the other hand, when $K>K_c$, $y$ is non-zero but given by 
Eq.~(\ref{ysol}), and one can see that 
$\Sigma(y)\simeq 1$, $\Sigma'(y)\simeq -1$ and 
\begin{equation}
\Sigma''(y)\simeq\left\{
\begin{array}{ll}
\frac{1}{2K_c}-\frac{\Delta}{K_c}&{\rm (I)},\\
\frac{1}{2K_c}-\frac{1-\mu}{2\mu}\frac{\Delta}{K_c}&{\rm (II)}.
\end{array}
\right.
\label{sigmaderiv1}
\end{equation}
From these relations, 
the mean cluster size for $K$ around $K_c$ is obtained as 
\begin{equation}
\bar{s}\simeq 
\left\{\begin{array}{ll}
  {c_-\over (-\Delta)}&(\Delta<0)\\
  {c_+\over \Delta}&(\Delta>0)
\end{array}\right. ,
\label{meancluster}
\end{equation}
where
\begin{equation}
c_-=2K_c, \quad 
c_+=
\left\{
        \begin{array}{ll}
        2K_c&{\rm (I)},\\
        {2\mu\over 1-2\mu} K_c &{\rm (II)}.
        \end{array}
\right.
\end{equation}
$\bar{s}$ diverges at $K_c$ both for (I) and (II). 
Thus, if we define $\chi \sim |\Delta|^{-\gamma}$, then 
$\gamma=1$ for both (I) and (II).

For (III),   $y$, the solution of  Eq.~(\ref{ysc}) is zero only when 
$K$ is zero. We suppose that $K$ is non-zero but small in the 
thermodynamic limit $N\to\infty$. Then, from Eq.~(\ref{sigma1b}), 
it follows that 
\begin{eqnarray}
\Sigma_1(y)&\simeq& 
1-\left({2B\over \mu}\right)^{\mu\over 2\mu-1} K^{\mu\over
        2\mu-1}+\left({2\over \mu}\right)^{1\over 2\mu-1}
        B^{2\mu\over 2\mu-1} K^{1\over 2\mu-1},
        \nonumber \\
\Sigma_1'(y)&\simeq&
-1+{1\over 2} \left({2B\over \mu}\right)^{\mu\over
        2\mu-1} K^{1-\mu\over 2\mu-1},
\nonumber \\
\Sigma_1''(y)&\simeq&
\frac{1-\mu}{2\mu K},
\end{eqnarray}
where $B=(1-\mu)^{1/\mu}|\Gamma(1-1/\mu)|$. 
Then the  mean cluster size is 
\begin{eqnarray}
\bar{s}&\simeq& 1-
\left({2B\over \mu}\right)^{\mu\over 2\mu-1} K^{\mu\over
        2\mu-1}+\left({2\over \mu}\right)^{1\over 2\mu-1}
        B^{2\mu\over 2\mu-1} K^{1\over 2\mu-1}+{2\mu K \over 2\mu-1} 
\left(1-{1\over 2} \left({2B\over \mu}\right)^{\mu\over
        2\mu-1} K^{1-\mu\over 2\mu-1}\right)^2\nonumber \\
&\simeq&
1+{2\mu \over 2\mu-1}K -{4\mu-1 \over 2\mu-1}
\left({2B \over \mu}\right)^{\mu\over 2\mu-1} K^{\mu \over
        2\mu-1}+{\mu^2 \over 2\mu-1}\left({2B\over \mu}\right)^{2\mu\over
        2\mu-1}K^{1\over 2\mu-1},
\end{eqnarray}
for small $K$. 
As shown in the 
numerical solutions for the mean cluster size  $\bar{s}$ 
obtained from Eqs.~(\ref{chi3}) and (\ref{ysc}) plotted in 
Fig.~\ref{fig:num_sol}, 
the most important feature of $\bar{s}$ for (III) is that it does not 
diverge at any value of $K$ but has only a finite peak at $K_{p2} =\mathcal{O}(1)$. 
It implies that there is no phase transition for (III), i.e., $2<\lambda<3$. 

\subsection{Number of loops and clusters}
The number of loops per vertex $\langle N_{\rm loop}\rangle/N$, 
which we denote by $\ell$, is also represented in terms of $\Sigma_1(y)$ 
as 
\begin{equation}
\ell=-F(y,0)=-1+y-{1\over 4K}y^2+\Sigma_1(y),
\label{ell2}
\end{equation}
with $y$ being the solution of Eq.~(\ref{ysc}). 

When $K<K_c$ for (I) and (II), and $K=0$ for (III), 
the value of $y$ is zero and $\Sigma_1(0)=1$, which leads to 
\begin{equation}
\ell=0.
\end{equation}
On the other hand, when $K>K_c$ for (I, II),  or 
$K>0$ for (III), the value of $\ell$ is not zero.   
From the behavior of $\Sigma_1(y)$ for small $y$ and Eq.~(\ref{ysol}), 
one can see that for $\Delta>0$ (I, II) or $K>0$ (III),
\begin{equation}
\ell\simeq\left\{
\begin{array}{ll}
{2\over 3}{(1-3\mu)^2 \over (1-2\mu)^3}\Delta^3 
& ({\rm I}),\\[2.5mm]
{1\over 2}\left({\mu \over (1-2\mu)\Gamma\left(1-\frac{1}{\mu}\right)}
                \right)^{2\mu \over 1-2\mu}\Delta^{1\over 1-2\mu}&({\rm II}),
\\[2.5mm]
{2\mu-1\over 4}
\left({2(1-\mu)^{1\over \mu}\left|\Gamma\left(1-\frac{1}{\mu}\right)\right|\over
                \mu}\right)^{2\mu\over 2\mu-1} K^{1\over 2\mu-1}&
({\rm III}).
\end{array}
\right.
\label{ell3}
\end{equation}
The exact solutions for $\ell$ are shown in Fig.~\ref{fig:num_sol}.
The number of clusters is simply related to $\ell$ as 
$\langle C \rangle /N = 1 - K +\ell $. 

\section{Cluster size distribution and largest cluster size}
\label{sec:distribution}

Beyond the largest cluster size or the mean cluster size, 
the whole distribution of cluster size $P(s)$ for the static model
can be derived  
from Eqs.~(\ref{sp3}) and (\ref{mag2}), which 
gives the parametrized equations for $\mathcal{P}(z)=1-m(1,h_0=-\ln z)$ as    
\begin{eqnarray}
&&z={1-{y\over 2K}\over \sum_{i=1}^N P_i e^{-NP_iy}}
=-{1-{y\over 2K}\over \Sigma_1'(y)},\nonumber \\
&&\mathcal{P}(z)={z\over N}\sum_{i=1}^N e^{-NP_iy}=z\Sigma_1(y),
\label{pz}
\end{eqnarray}
where $\Sigma_1(y)$ in Appendix A is used. 
$P(s)$ is obtained by 
$P(s)=(1/s!) (d^s/dz^s) \mathcal{P}(z)|_{z=0}$. 
In particular, $P(s)$ for $s\gg 1$ is 
contributed to by such a singular term as $(z_0-z)^x$ 
with $x$ a non-integer in $\mathcal{P}(z)$.
The functional form of $\mathcal{P}(z)$ depends on $P_i$ 
for $1\leq i\le N$. 
In this section, we solve Eqs.~(\ref{sp3}) and (\ref{mag2}) 
when $P_i$ is given by Eq.~(\ref{Pi}) to 
find the cluster size distribution $P(s)$. 
Furthermore, we derive the largest cluster size $\langle S\rangle$ 
before a giant cluster appears 
through the following relation~\cite{cohen_book} 
\begin{equation}
\sum_{s\ne \langle S\rangle}P(s)=1-{\langle S \rangle \over N},
\label{sumrule}
\end{equation}
which is equivalent to the relation $m(1,h_0\to 0) =\langle S\rangle/N$ 
in the limit $h_0 \langle S\rangle\to\infty$ in Eq.~(\ref{m10}). 

\subsection{$\mu=0$ : Erd\H{o}s-R\'{e}nyi model}
Before considering the case of $0<\mu<1$ in Eq.~(\ref{Pi}), 
we first consider the Erd\H{o}s-R\'{e}nyi model with $P_i=1/N$ 
corresponding to the case of $\mu=0$. 
In this case, the parametrized equations for $\mathcal{P}(z)$, 
Eq.~(\ref{pz}) is simply written as 
\begin{eqnarray}
&&z=\left(1-{y\over 2K}\right)e^y, \nonumber \\
&&\mathcal{P}(z)=ze^{-y}.
\end{eqnarray}
If we consider $y$ as a function of $z$, 
it has the properties $y(z=0)=2K$ and $dy/dz<0$ for  $0\leq z<z_0\equiv e^{2K-1}/2K$. 
On the other hand, $\mathcal{P}(z=0)=0$ and 
$\mathcal{P}(z)$ is an increasing function of $z$. 
By substituting $z=1$, we find that the giant cluster size $m=1-\mathcal{P}(1)$ is 
non-zero for $K>K_c=1/2$, and especially, $m$ is given by $m\simeq 2\Delta$  
for $0<\Delta=K/K_c-1\ll 1$.

Around $z_0$, the function $y(z)$ becomes singular as 
$y(z)\simeq 2K-1+[4Ke^{1-2K}]^{1/2}(z_0-z)^{1/2}$. 
Also, $\mathcal{P}(z)$ has the square-root singularity at $z_0$ as 
\begin{equation}
\mathcal{P}(z)\simeq {1\over 2K}-\left({e^{1-2K}\over K}\right)^{1\over 2}
(z_0-z)^{1\over 2}.
\end{equation}
Differentiating $\mathcal{P}(z)$ at $z=0$, one can 
obtain $P(s)$, which is given for large $s$ as 
\begin{eqnarray}
P(s)&=& \left.{1\over s!}{d^s \over dz^s} \mathcal{P}(z)\right|_{z=0}\nonumber\\
&\simeq& 
-\left({e^{1-2K}\over K}\right)^{1\over 2}z_0^{-s+\frac{1}{2}} 
{\Gamma\left(s-\frac{1}{2}\right)\over \Gamma(s+1)\Gamma\left(-\frac{1}{2}\right)}  
\nonumber \\
&\simeq& \left({e^{1-2K+\frac{1}{s_0}}\over 4\pi K}\right)^{1\over2} 
s^{-\frac{3}{2}}e^{-\frac{s}{s_0}},
\label{Ps0}
\end{eqnarray}
where $\Gamma(-1/2)=-2\pi^{1/2}$ is used and $s_0\equiv 1/\ln z_0$. 
One can notice that $z_0=1$, $s_0\to \infty$, and $P(s)\sim s^{-3/2}$   
at $K_c$. When $|\Delta|\ll 1$, 
the cut-off $s_0$ is approximately $2/\Delta^2$.

The presence of the cut-off $s_0$  means that a cluster 
of size $s$ larger than $s_0$ can be found only with the exponentially small 
probability $\sim e^{-s/s_0}$. Thus, the largest cluster 
is as large as $s_0$ before a giant cluster appears. 
$\langle S\rangle$ increase as $K$ approaches $K_c$. 
However, in finite size systems, the largest clusters 
cannot grow infinitely as $K\to K_c$, which 
is obvious from Eq.~(\ref{sumrule}). 
Suppose that $\langle S\rangle$ is much less than $s_0$. Then, 
one can easily see that $S\sim N^{2/3}$ applying Eq.~(\ref{Ps0}) to 
Eq.~(\ref{sumrule}). It indicates that in the regime of $K$ where 
$s_0\gg N^{2/3}$, or $-1\ll \Delta N^{1/3}<0$, 
$\langle S\rangle$ is $\mathcal{O}(N^{2/3})$. 
For $K>K_c$, the largest cluster size in the finite size system is given by 
$Nm\simeq 2N\Delta$ only when $N\Delta\gg N^{2/3}$. 
To summarize, $\langle S\rangle$ is given by 
\begin{equation}
\langle S\rangle\sim \left\{
\begin{array}{ll}
\frac{1}{\Delta^{2}}&(\Delta N^{1\over 3}\ll -1),\\
N^{2\over 3} &(|\Delta N^{1\over 3}|\ll 1),\\
N \Delta &(\Delta N^{1\over 3}\gg 1).
\end{array}
\right.
\label{largestcluster0}
\end{equation}
The regime of $K$ satisfying $|\Delta N^{1/3}|\ll 1$ in finite size systems 
shrinks to a point $K_c$ in the thermodynamic limit $N\to\infty$, which 
we will call the critical regime. 
If we introduce a scaling exponent $\bar{\nu}$ to describe 
the critical regime as $|\Delta N^{1/\bar{\nu}}|\ll 1$, 
$\bar{\nu}=3$ in the ER model.  

\subsection{The case (I): $0<\mu<1/3$}
As shown in Appendix A, $\Sigma_1(y)$ has a singular term with the 
$\mu$-dependent exponent in its expansion in $y$, which 
allows $\mathcal{P}(z)$ to have singularity other than 
the square-root one for the ER model. 

The first relation of Eq.~(\ref{pz}) is expanded in $y$ as, using Eq.~(\ref{sigma1o}),  
\begin{equation}
z=\sum_{n=0}^\infty a_ny^n,
\label{zregular}
\end{equation}
where the first few coefficients are 
\begin{eqnarray}
a_0&=&1,\nonumber \\
a_1&=&{1\over 2K_c}-{1\over 2K}={\Delta\over 2K},\nonumber \\
a_2&=&- \left[{1\over 2}(1-\mu)^3 N^{3\mu-1}\zeta_N(3\mu) - 
{\Delta\over 4KK_c}\right].
\label{coefficients}
\end{eqnarray}
Here, the critical point $K_c(N)$ is given by 
\begin{equation}
K_c(N)={1\over 2(1-\mu)^2N^{2\mu-1}\zeta_N(2\mu)}.
\label{kcn}
\end{equation}
When $0\leq \mu<1/2$, 
the solution $y$ of Eq.~(\ref{zregular}) with $z=1$ is $0$ 
and thus $\mathcal{P}(1)=1$ for $a_1<0$ while 
$y$ is a positive value and $\mathcal{P}(1)<1$ for $a_1>0$. 
Therefore, $K_c(N)$ is the percolation transition point and 
indeed, 
converges to the $K_c$ in Eq.~(\ref{kc}) in the thermodynamic limit 
$N\to\infty$. 
However, when $1/2<\mu<1$, $K_c(N)=\mathcal{O}(N^{-(2\mu-1)})$ and 
$y$ satisfying Eq.~(\ref{zregular}) with $z=1$ is $\sim N^{-\mu}\Delta$ 
for $K>K_c(N)$, which goes to zero in the thermodynamic limit.  
It means that $K_c(N)$ is not the percolation transition point for $1/2<\mu<1$. 

When $y\gg N^{-\mu}/(1-\mu)$, 
Eq.~(\ref{sigma1b}) should be used  and thus 
\begin{equation}
z=1-{y\over 2K}+\sum_{n=1}^{\lfloor \frac{1}{\mu}-1\rfloor}a'_n y^n
+Ay^{\frac{1}{\mu}-1}+\cdots, 
\label{zsingular}
\end{equation}
where $a'_1=1/(2K_c)$, $a'_n=a_n$ for $n\ge 2$, and 
$A=\Gamma(2-1/\mu) (1-\mu)^{1/\mu-1}$. 

Similarly to the ER model,  when $0<\mu<1/3$, 
the value of $m=1-\mathcal{P}(1)$ is nonzero for  $K>K_c=\mathcal{O}(1)$, 
and thus $K_c$ can be identified with the percolation transition point. 

The leading singular term of the function $y(z)$ varies depending on $z$. 
First, when $1-s_m^{-1}\ll z<z_0=1+s_0^{-1}$  with 
\begin{eqnarray}
s_0&\equiv&{16K^2 |a_2|\over \Delta^2},\nonumber \\
s_m&\equiv&{2K(1-\mu)^2 N^{2\mu}\over 
{2K|a_2|-\Delta (1-\mu)N^\mu}} \quad 
\left(\Delta<{4K|a_2|\over (1-\mu)N^\mu}\right),   
\label{s0sm}
\end{eqnarray}
the function $y(z)$ is represented as, from Eq.~(\ref{zregular}),  
\begin{equation}
y(z)\simeq {\Delta \over 4K|a_2|} 
+ {(z_0-z)^{1\over 2}\over |a_2|^{1\over 2}}.
\label{yz1a}
\end{equation}
Notice that $y(z)$ with $z\gg 1-s_m^{-1}$ satisfies 
the relation $y(z)\ll N^{-\mu}/(1-\mu)$.  
Next, when $z\ll 1-s_m^{-1}$, $y(z)$ is expanded from Eq.~(\ref{zsingular})  
as
\begin{equation}
y(z)\simeq {\Delta \over 4K|a_2|} 
+|a_2|^{-\frac{1}{2}} \left[\phantom{1\over 2}z_0-z+\cdots
+ A \left({\Delta \over 4K |a_2|} +
{(z_0-z)^{1\over 2}\over |a_2|^{1\over 2}}\right)^{\frac{1}{\mu}-1}\right]^{1\over 2}. 
\label{yz1b}
\end{equation}
This implies that $y(z)$ has the square-root singularity 
as in Eq.~(\ref{yz1a}) except for the case of $\Delta<0$ and 
$1-s_0^{-1}\ll z\ll 1-s_m^{-1}$,  where $y(z)$ is given by 
\begin{equation}
y(z)\simeq {2K\over |\Delta|} (1-z)+ \cdots 
+A \left({2K\over |\Delta|}\right)^{1\over \mu} (1-z)^{\frac{1}{\mu}-1}+\cdots.
\label{yz1c}
\end{equation}
Such regime of $z$ exists when $s_0\ll s_m$.

Different singularities of $y(z)$ depending on the range of $z$ and 
$\Delta$ shown in Eqs.~(\ref{yz1a}), (\ref{yz1b}), and (\ref{yz1c}) 
are inherited to $\mathcal{P}(z)$ by the relation 
$\mathcal{P}(z)=z\Sigma_1(y)$, 
and in turn, cause $P(s)$ to behave distinctively depending on 
the range of $s$  and $\Delta$. 
When $\Delta\ll -N^{-\mu}$, $s_0$ is much less than $s_m$ and 
\begin{equation}
P(s)\sim \left\{
\begin{array}{ll}
s^{-\frac{3}{2}} e^{-\frac{s}{s_0}} & (1\ll s\ll s_0), \\
(|\Delta|s)^{-\frac{1}{\mu}} & (s_0\ll s\ll s_m), \\
s^{-\frac{3}{2}}e^{-\frac{s}{s_0}}&(s\gg s_m).
\end{array}
\right.
\label{Ps1a}
\end{equation}
$P(s)$ for $1\ll s\ll s_0$ is related to Eq.~(\ref{yz1b}), 
that for $s_0\ll s\ll s_m$ to Eq.~(\ref{yz1c}), and 
that for $s\gg s_m$ to Eq.~(\ref{yz1a}). 
On the other hand, 
when $\Delta\gg -N^{-\mu}$,   the regime of 
$z$ where $y(z)$ follows Eq.~(\ref{yz1c}) vanishes but 
$P(s)$ is  simply given by 
\begin{equation}
P(s)\simeq 
s^{-\frac{3}{2}} e^{-\frac{s}{s_0}} 
\label{Ps1b}
\end{equation}
for all $s$.

When $\Delta\ll -N^{-\mu}$, the cluster size distribution 
decays exponentially for $s\gg s_m$, so the largest 
cluster is as large as $s_m\sim N^\mu/|\Delta|$. 
When $\Delta\gg -N^{-\mu}$, the cluster size distribution 
takes the same form as that for the ER model, and 
the critical regime is specified with the same exponent 
$\bar{\nu}=3$. 
Consequently,  
\begin{equation}
\langle S\rangle\sim \left\{
\begin{array}{ll}
\frac{N^\mu}{|\Delta|}&(\Delta \ll -N^{-\mu}),\\
\frac{1}{|\Delta|^2}&(-N^{-\mu}\ll \Delta\ll -N^{-\frac{1}{3}}), \\
N^{2\over 3} &(|\Delta N^{1\over 3}|\ll 1),\\
N \Delta &(\Delta N^{1\over 3}\gg 1),
\end{array}
\right.
\label{largestcluster1}
\end{equation}
where Eq.~(\ref{ysol}) is used to obtain $\langle S \rangle$ for $\Delta\gg N^{-1/3}$. 

\subsection{The case (II): $1/3<\mu<1/2$}
Eqs.~(\ref{zregular}) and (\ref{zsingular}) 
are valid also for $1/3<\mu<1/2$. 
However, one should note that the singular term
$y^{1/\mu-1}$ is the next leading term to $y$ in Eq.~(\ref{zsingular}), 
which causes $\mathcal{P}(z)$ to have an $\mu$-dependent 
singularity other than the square-root one even in 
the critical regime. 

Now we consider the singularity of the function $y(z)$. 
As for $0<\mu<1/3$, when $1-s_m^{-1}\ll z\ll z_0=1+s_0^{-1}$ 
with $s_0$ and $s_m$ in Eq.~(\ref{s0sm}), $y(z)$ is given by Eq.~(\ref{yz1a}).
Note that $a_2=\mathcal{O}(N^{3\mu-1})$ for $\mu>1/3$. 
$y(z)$ for $z\ll 1-s_m^{-1}$  is obtained by inverting Eq.~(\ref{zsingular}). 
When $\Delta>0$, if the regime $1-s_c^{-1}\ll z\ll 1-s_m^{-1}$ with  
\begin{equation}
s_c\equiv {1-\mu\over 1-2\mu}
\left(|A|\frac{1-\mu}{\mu}\right)^{\mu\over 1-2\mu} 
|\Delta|^{-{1-\mu\over 1-2\mu}},
\label{sc2}
\end{equation}  
exists,   
$y(z)$ in that regime is  given as 
\begin{eqnarray}
y(z)&\simeq& 
\left({\mu \Delta\over 2K|A|(1-\mu)}\right)^{\mu\over 1-2\mu}
\nonumber \\
&+&     \left[{2\mu\over 1-2\mu} 
\left({\Delta\over 2K}\right)^{3\mu-1\over 1-2\mu}  
        \left(|A|\frac{1-\mu}{\mu}\right)^{-{\mu\over 1-2\mu}}
                \right]^{1\over 2}(z_c-z)^{1\over 2}\nonumber \\
		&+&\cdots,
\label{yz2a}
\end{eqnarray}
with $z_c=1+s_c^{-1}$. 
When $\Delta<0$ and  $1-s_c^{-1}\ll z\ll 1-s_m^{-1}$, 
$y(z)$ is given by 
\begin{equation}
y(z)\simeq {2K\over |\Delta|}(1-z)-|A|
\left({2K\over |\Delta|}\right)^{1\over \mu}
(1-z)^{\frac{1}{\mu}-1}+\cdots.
\label{yz2b}
\end{equation}
Finally, both for $\Delta>0$ and $\Delta<0$, 
if $z$ is in the regime $z\ll 1- \max\{s_c^{-1},s_m^{-1}\}$,   
$y(z)$ exhibits the following singularity as 
\begin{equation}
y(z)\simeq 
\left({1-z\over |A|}\right)^{\mu\over 1-\mu}+\cdots.
\label{yz2c}
\end{equation}

The functional form of the cluster size distribution 
varies depending on $s$ and $\Delta$. 
When $\Delta\ll -N^{-(1-2\mu)}$, $\mathcal{P}(z)$ has 
three different singularities, Eqs.~(\ref{yz1a}), (\ref{yz2b}), 
and (\ref{yz2c}), in the corresponding regimes of $z$, 
and thus $P(s)$ is given as 
\begin{equation}
P(s)\simeq \left\{
\begin{array}{ll}
s^{-{1\over 1-\mu}} & (1\ll s\ll s_c), \\
(|\Delta|s)^{-\frac{1}{\mu}} & (s_c\ll s\ll s_m), \\
s^{-\frac{3}{2}}e^{-\frac{s}{s_0}}&(s\gg s_m).
\end{array}
\right.
\label{Ps2a}
\end{equation}
When $-N^{-(1-2\mu)}\ll \Delta \ll N^{-(1-2\mu)}$,  
$\mathcal{P}(z)$ is contributed to by Eqs.~(\ref{yz1a}) and 
(\ref{yz2c}), which leads to 
\begin{equation}
P(s)\simeq \left\{
\begin{array}{ll}
s^{-{1\over 1-\mu}} & (1\ll s\ll s_m), \\
s^{-\frac{3}{2}}e^{-\frac{s}{s_0}}&(s\gg s_m).
\end{array}
\right.
\label{Ps2b}
\end{equation}
Finally, when $\Delta\gg N^{-(1-2\mu)}$, 
the regime of $z$ where Eq.~(\ref{yz1a}) is valid disappears, and 
\begin{equation}
P(s)\simeq \left\{
\begin{array}{ll}
s^{-{1\over 1-\mu}} & (1\ll s\ll s_c), \\
s^{-\frac{3}{2}}e^{-\frac{s}{s_c}}&(s\gg s_c),
\end{array}
\right.
\label{Ps2c}
\end{equation}
as Eqs.~(\ref{yz2a}) and (\ref{yz2c}) imply.  

The largest cluster size follows $s_m$ up to $\Delta\sim -N^{-(1-2\mu)}$ 
beyond which $\langle S\rangle$  is $\mathcal{O}(N^{1-\mu})$ 
as shown by Eq.~(\ref{sumrule}). 
The comparison of $N^{1-\mu}$ and 
the giant cluster size $mN\sim N\Delta^{\mu/(1-2\mu)}$ given in
Eq.~(\ref{ysol}) indicates that the largest cluster size 
is given by the latter when $\Delta\gg N^{-(1-2\mu)}$. 
In summary, 
the largest cluster size is  
\begin{equation}
\langle S\rangle\sim \left\{
\begin{array}{ll}
\frac{N^\mu}{|\Delta|}&(\Delta N^{1-2\mu}\ll -1),\\
N^{1-\mu} &(|\Delta N^{1-2\mu}|\ll 1),\\
N \Delta^{\mu\over 1-2\mu} &(\Delta N^{1-2\mu}\gg 1),
\end{array}
\right.
\label{largestcluster2}
\end{equation}
and the exponent $\bar{\nu}$ is $1/(1-2\mu)$ for $1/3<\mu<1/2$.

\subsection{The case (III): $1/2<\mu<1$}
For $1/2<\mu<1$,
the value of $K_c(N)$ given in Eq.~(\ref{kcn}) 
is $\mathcal{O}(N^{-(2\mu-1)})$ and thus,  
for all finite $K>0$, the giant cluster  size 
$m=1-\mathcal{P}(1)$ is 
non-zero in the thermodynamic limit $N\to\infty$. 
However, $m$ is given as $\sim K^{\mu/(2\mu-1)}$, 
which is $\mathcal{O}(N^{-\mu})$ around $K_c(N)$ and 
vanishes in the thermodynamic limit $N\to\infty$. 
We consider finite size systems where $K_c(N)$ is non-zero but finite 
and investigate how the cluster size distribution and the 
largest cluster size behave around $K_c(N)$. 

Following the same step as for $0<\mu<1/2$, 
one finds that when $1-s_m^{-1}\ll z\ll 1+s_0^{-1}$ 
with $s_0$ and $s_m$ in Eq.~(\ref{s0sm}), 
$y(z)$ is given by Eq.~(\ref{yz1a}).
Eq.~(\ref{zsingular}) applies  to the case of $z\ll 1-s_m^{-1}$. 
If the regime $1-s_c^{-1}\ll z\ll 1-s_m^{-1}$ with 
\begin{equation}
s_c\equiv {1-\mu\over 2\mu-1} \left(A\frac{1-\mu}{\mu}\right)^{-\frac{\mu}{2\mu-1}}
K^{-{1-\mu\over 2\mu-1}}, 
\label{sc3}
\end{equation}
exists, $y(z)$ is given by 
\begin{equation}
y(z)\simeq \left(2KA\frac{1-\mu}{\mu}\right)^{\mu\over 2\mu-1}
+\left[{2\mu\over 2\mu-1} (2K)^{3\mu-1\over 2\mu-1} 
\left(A\frac{1-\mu}{\mu}\right)^{\mu\over 2\mu-1}\right]^{1\over 2}
(z_c-z)^{1\over 2}+\cdots,
\label{yz3a}
\end{equation}
with $z_c=1+s_c^{-1}$. 
In the regime $z\ll 1-\max\{s_c^{-1}, s_m^{-1}\}$,  $y(z)$ is  
given by 
\begin{equation}
y(z)\simeq 2K(1-z)+A(2K)^{1\over \mu} (1-z)^{\frac{1}{\mu}-1}+\cdots.
\label{yz3b}
\end{equation}

When $K\lesssim K_c(N)$,  the generating function 
$\mathcal{P}(z)$ is evaluated by Eq.~(\ref{yz1a}) and Eq.~(\ref{yz3b}) 
to give the cluster size distribution $P(s)$ as follows:
\begin{equation}
P(s)\simeq \left\{
\begin{array}{ll}
\left(\frac{s}{K}\right)^{-{1\over \mu}} & (1\ll s\ll s_m), \\
N^{-{3\mu-1\over 2}} s^{-\frac{3}{2}}e^{-\frac{s}{s_0}}&(s\gg s_m).
\end{array}
\right.
\label{Ps3a}
\end{equation}
where the factor $N^{-(3\mu-1)/2}$ comes from $|a_2|^{-1/2}$. 
Eq.~(\ref{Ps3a}) is valid when $s_m<s_c$. 
When $K\gg  K_c(N)$,  the regime of $z$ where Eq.~(\ref{yz1a}) applies  
disappears and  $\mathcal{P}(z)$ is evaluated using 
Eqs.~(\ref{yz3a}) and (\ref{yz3b}). Consequently,  
$P(s)$ is 
\begin{equation}
P(s)\simeq \left\{
\begin{array}{ll}
\left(\frac{s}{K}\right)^{-{1\over \mu}} & (1\ll s\ll s_c), \\
K^{3\mu-1\over 2(2\mu-1)}s^{-\frac{3}{2}}e^{-\frac{s}{s_c}}&(s\gg s_c),
\end{array}
\right.
\label{Ps3b}
\end{equation}
for $K\gg K_c(N)$. 

The largest cluster is as large as $s_m$ approximately up to $K_c(N)$
and is given by $mN$ in Eq.~(\ref{ysol}) for $K\gg K_c(N)$. That is,   
\begin{equation}
\langle S\rangle\sim \left\{
\begin{array}{ll}
KN^\mu &(K\lesssim K_c(N)),\\
N K^{\mu\over 2\mu-1} &(K_c(N)\ll K\ll 1).
\end{array}
\right.
\label{largestcluster3}
\end{equation}
This result implies that the largest cluster size 
is still $\mathcal{O}(N^{1-\mu})$ even when $K>K_c(N)$ 
unless $K\gg K_c(N)$, consistent with Eq.~(\ref{ysol}). 
The conditions $K\lesssim K_c(N)$ and $K\gg K_c(N)$ 
can be rewritten as $\Delta\lesssim 1$ and $\Delta\gg 1$ 
with $\Delta=K/K_c(N)-1$, and thus the exponent $\bar{\nu}$ to 
define the critical regime is infinity for $1/2<\mu<1$. 
The absence of a critical regime with a finite $\bar{\nu}$ means 
that there is no percolation transition in the evolution 
of scale-free graphs with $2<\lambda<3$, as the absence of 
a divergence in the mean cluster size does. 

\section{Numerical simulations and finite size scaling}
\label{sec:fss}

\begin{figure}[t]
\centering
\includegraphics[width=8.9cm]{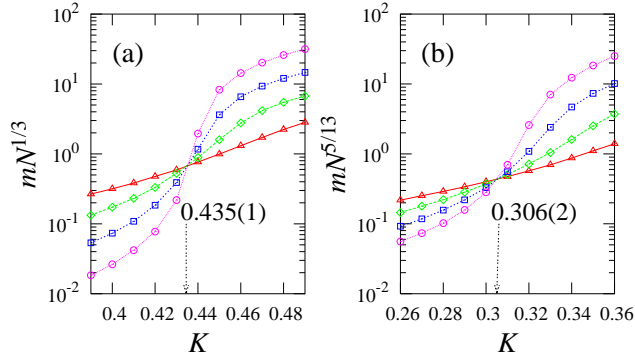}
\caption{Plots of $mN^{\beta/\bar{\nu}}$ vs. $K$ 
for $\mu=5/19$ (a) and $\mu=5/13$  (b). 
The number of vertices $N$ is 
$10^4(\triangle)$, $10^5 (\Diamond)$, $10^6 (\square)$, and $10^7 (\bigcirc)$. 
The data cross at $K_c=0.435(1)$ (a) and $K_c=0.306(2)$ (b), respectively, 
which are in accordance with $0.436$ and $0.305$ from Eq.~(\ref{kc}).}
\label{fig:crossing}
\end{figure}
\begin{figure*}
\centering
\includegraphics[width=14cm]{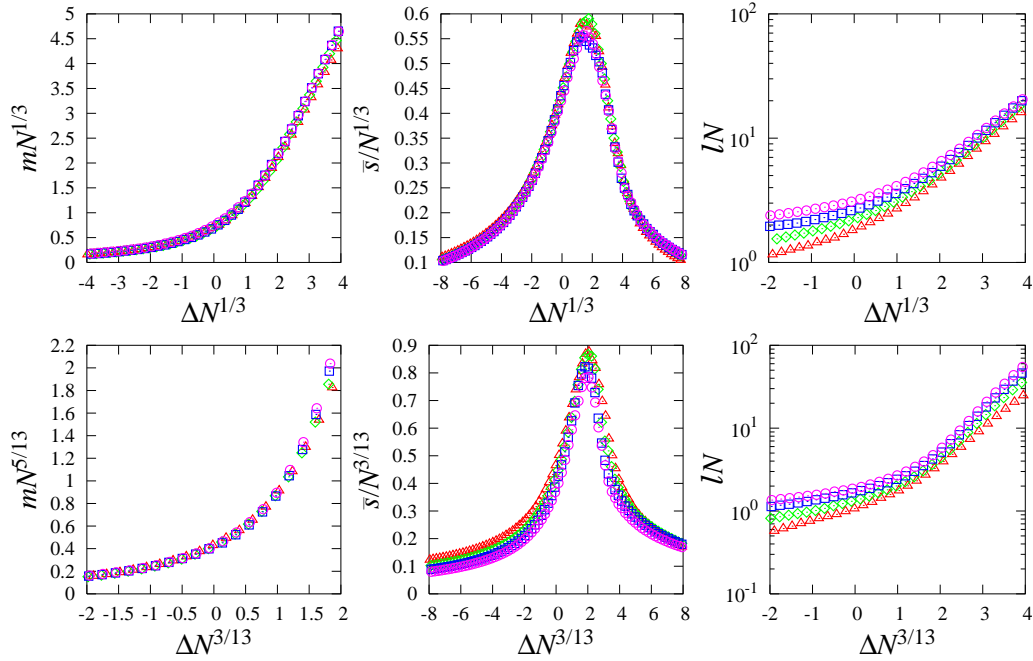}
\caption{Data collapse of scaled 
        giant cluster sizes $mN^{\beta/\bar{\nu}}$,  
mean cluster sizes $\bar{s}/ N^{1/\bar{\nu}}$,  and 
number of loops $\ell N$ plotted vs. $\Delta N^{1/\bar{\nu}}$ for $\mu=5/19$
(upper) and $\mu=5/13$  (lower). $N$ is $10^4(\triangle)$, $10^5 (\Diamond)$, $10^6 (\square)$, and $10^7 (\bigcirc)$.}
\label{fig:m&x}
\end{figure*}

\begin{figure}
\centering
\includegraphics[width=8.9cm]{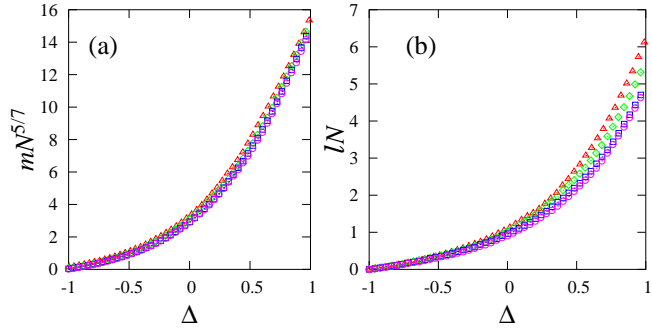}
\caption{Data collapse  of the scaled giant cluster size (a) and 
number of loops (b) vs. $\Delta$ for $\mu=5/7$.
$N$ is  $10^4(\triangle)$, $10^5 (\Diamond)$, $10^6 (\square)$, 
and $10^7 (\bigcirc)$.}
\label{fig:ml24}
\end{figure}

\begin{figure}
\centering
\includegraphics[width=8.9cm]{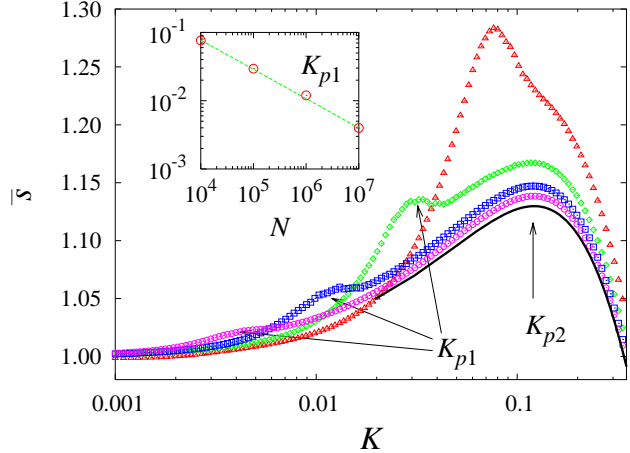}
\caption{
Mean cluster size $\bar{s}$ 
as a function of $K$ in semi-logarithmic scales 
with $\mu=5/7$ 
for $N=10^4(\triangle)$, $10^5 (\Diamond)$, $10^6 (\square)$,  and 
$10^7 (\bigcirc)$.
In addition to the peak at $K_{p1}$, another peak is shown at 
$K_{p2}\simeq 0.1 \, N$ for  $N=10^5$, $10^6$, and $10^7$, respectively. 
The solid line represents the exact solution obtained from 
Eqs.~(\ref{ysc}) and (\ref{chi3}). The measured values of $K_{p1}$ 
($\bigcirc$) as a function of $N$ are plotted in the inset 
together with the guide line whose slope is $1-2\mu$ 
for comparison.} 
\label{fig:x24}
\end{figure}

In this section, we derive the finite-size scaling forms 
of the giant cluster size $m$, the mean cluster size $\bar{s}$, 
and the number of loops $\ell$ and check against their numerical data 
from simulations for the static model. 
Investigating the cluster size distribution and 
the largest cluster size, 
we identified the critical regimes and the corresponding scaling 
variables, which enables us to predict the scaling behaviors 
of $m$, $\bar{s}$, and $\ell$. 

Let us sketch the numerical procedure briefly.
At each step, we pick two vertices $i$ and $j$,
with probabilities $P_i$ and $P_j$, respectively.
Then we put an edge between $i$ and $j$ unless there is
one already. Repeat the procedure until there are $L$ edges
made in the system. 
The generation of random numbers with non-uniform probability
density $P_i$ is the most time-consuming part here.
One efficient way is to use, e.g.,             
Walker's algorithm \cite{walker} combined with the so-called
Robin Hood method \cite{robinhood}, which is explained in Appendix B.
This method is exact and takes time of ${\mathcal O}(N)$ to set up a table
and ${\mathcal O}(1)$ to choose a vertex, 
making a large-size simulation feasible. 
Once a graph is constructed, the clusters are identified by the 
standard breadth-first search, 
during which we can extract 
the size of the largest cluster, 
the mean cluster size excluding the largest 
one as defined in Eq.~(\ref{chi}), 
and the number of clusters within the graph.
Recall that the number of clusters is related to 
the number of loops by Eq.~(\ref{lnc}). 
We choose three values of $\mu$, 
$\mu=5/19$ ($\lambda=4.8$), $5/13$ ($\lambda=3.6$), 
and $5/7$ ($\lambda=2.4$) as representatives of the regime 
(I), (II), and (III), respectively.
We perform the simulations for the system size $N$ up to $10^7$, 
and the ensemble average is evaluated with at least $10^3$ runs.

Now we first consider the regime $0<\mu<1/2$ or $\lambda>3$. 
The scaling behaviors of the giant cluster size $m$ in this regime 
are shown in Eqs.~(\ref{largestcluster1}) and (\ref{largestcluster2}) 
and can be written as 
\begin{equation}
m=N^{-\frac{\beta}{\bar{\nu}}} \Psi_{\rm (I,II)}(\Delta N^{1\over \bar{\nu}}), 
\label{scaling:m}
\end{equation}  
where the scaling exponents $\beta$ and $\bar{\nu}$ are given by 
\begin{eqnarray}
\beta&=&\left\{
\begin{array}{ll}
1 & (0<\mu<\frac{1}{3}),\\
\frac{\mu}{1-2\mu} & (\frac{1}{3}<\mu<\frac{1}{2}),
\end{array}
\right.
\nonumber \\
\bar{\nu}&=&\left\{
\begin{array}{ll}
3 & (0<\mu<\frac{1}{3}),\\
\frac{1}{1-2\mu} & (\frac{1}{3}<\mu<\frac{1}{2}).
\end{array}
\right.
\label{betanu}
\end{eqnarray}
The scaling function $\Psi_{\rm (I,II)}(x)$ behaves as
\begin{equation}
\Psi_{\rm (I,II)}(x)\sim \left\{
\begin{array}{ll}
{\rm const.}&(x\ll 1),\\
x^\beta & (x\gg 1).
\end{array}
\right.
\end{equation}

The critical point $K_c$ can be found numerically 
by plotting $mN^{\beta/\bar{\nu}}$ vs. $K$ as shown in
Fig.~\ref{fig:crossing}, which are in good agreement with 
Eq.~(\ref{kc}). 
In Fig.~\ref{fig:m&x}, the plots of $mN^{\beta/\bar{\nu}}$ vs.
$\Delta N^{1/\bar{\nu}}$ are shown for 
$\mu=5/19$ and $5/13$, respectively. The data collapse confirming 
the scaling behaviors in Eq.~(\ref{scaling:m}). 

In the critical regime, the cluster size distribution 
follows the power law $P(s)\sim s^{-\tau}$ with 
$\tau=3/2$ for (I) and $1/(1-\mu)$ for (II). 
Then the mean cluster size in the critical regime is 
given by $\bar{s}\sim \langle S\rangle^{2-\tau}$ since 
$\bar{s}=\sum_{s\ne \langle S\rangle} sP(s)$. 
Since $\langle S\rangle\sim N^{1-\beta/\bar{\nu}}$ in 
the critical regime as in Eq.~(\ref{scaling:m}), 
the mean cluster size is represented as 
$\bar{s}\sim N^{(1-\beta/\bar{\nu})(2-\tau)}
=N^{1/\bar{\nu}}$, which leads to, combined with Eq.~(\ref{meancluster}), 
\begin{equation}
\bar{s} = N^{1\over \bar{\nu}} \Phi_{\rm (I,II)}  (\Delta N^{1\over \bar{\nu}}),
\end{equation}
where the scaling function $\Phi_{\rm (I,II)}(x)$ behaves as 
\begin{equation}
\Phi_{\rm (I,II)}(x)\sim \left\{
\begin{array}{ll}
{\rm const.}&(x\ll 1),\\
x^{-1} & (x\gg 1).
\end{array}
\right.
\end{equation}
Plots of $\bar{s}/ N^{1/\bar{\nu}}$ vs. $\Delta N^{1/\bar{\nu}}$ 
for $\mu=5/19$ and $5/13$ are shown in Fig.~\ref{fig:m&x}. 

Since the giant cluster size and the mean cluster size 
correspond to the first and second derivative of the free energy 
with respect to the external field, respectively, 
it is natural that both have the same scaling variable. 
Therefore, the number of loops, corresponding to the free energy itself,  
is also described in terms of the 
scaling variable $\Delta N^{1/\bar{\nu}}$ with 
$\bar{\nu}$ in Eq.~(\ref{betanu}). Considering 
the number of loops for $K>K_c$ given in Eq.~(\ref{ell3}), 
one can see that 
\begin{equation}
\ell={1\over N} \Omega_{\rm (I,II)}(\Delta N^{1\over \bar{\nu}}),
\end{equation}
where the scaling function $\Omega_{\rm (I,II)}(x)$ behaves as 
\begin{equation}
\Omega_{\rm (I,II)}(x)\sim \left\{
\begin{array}{ll}
{\rm const.}&(x\ll 1),\\
x^{\bar{\nu}} & (x\gg 1).
\end{array}
\right.
\end{equation}
This implies that $\ell$ is $\mathcal{O}(N^{-1})$ 
in the critical regime, which 
is supported by the 
numerical simulation results in Fig.~\ref{fig:m&x}.

Next, we consider the regime $1/2<\mu<1$. 
From the largest cluster size shown in Eq.~(\ref{largestcluster3}),  
the giant cluster size $m$ can be written as 
\begin{equation}
m\sim N^{-\mu}~ \Psi_{\rm (III)}(\Delta),
\label{eq:S2}
\end{equation}
where the function $\Psi_{\rm (III)}(x)$ behaves as 
\begin{equation}
\Psi_{\rm (III)}(x)\sim \left\{
\begin{array}{ll}
{\rm const.}&(x\ll 1),\\
x^{\mu\over 2\mu-1} & (x\gg 1).
\end{array}
\right.
\end{equation}
Notice that $m$ increases smoothly passing $K_c(N)$ as manifested by 
$\Delta$ not scaling with $N$. 
Similarly, the number of loops $\ell$ 
represented in terms of the scaling variable $\Delta$ around $K_c(N)$  
should exhibit the following scaling behaviors to satisfy Eq.~(\ref{ell3}):
\begin{equation}
\ell={1\over N} \Omega_{\rm (III)}(\Delta),
\end{equation}
where the scaling function $\Omega_{\rm (III)}(x)$ behaves as 
\begin{equation}
\Omega_{\rm (III)}(x)\sim \left\{
\begin{array}{ll}
{\rm const.}&(x\ll 1),\\
x^{1\over 2\mu-1} & (x\gg 1).
\end{array}
\right.
\end{equation}
Data collapse of $mN^{\mu}$ and $\ell N$ vs. 
$\Delta$ is shown in Fig.~\ref{fig:ml24}.

The numerical data of the mean cluster size are shown in Fig.~\ref{fig:x24}.
As $N$ increases, the mean cluster size $\bar{s}$ approaches 
the exact solution in Eq.~(\ref{chi3}) represented by the solid line 
in Fig.~\ref{fig:x24}. It does not diverge at any value of $K$, but 
instead its peak height decreases as $N$ increases. 
We additionally find that 
the mean cluster size $\bar{s}$ has a small peak at 
$K_{p1}$, which scales as  $N^{1-2\mu}$ as shown in the inset 
of Fig.~\ref{fig:x24}. The value of $K_{p1}$ is not 
equal to $K_c(N)$ although they are the same order of $N$. 
The reason for the peak at $K_{p1}$ is as follows. 
As $K$ increases, 
the largest cluster size $\langle S\rangle$ 
and the mean cluster size $\bar{s}=\sum_{s\ne \langle S\rangle}sP(s)$ 
also increase.  However, as $K$ approaches $K_c(N)$, 
the cluster size distribution $P(s)$ begins to develop the 
exponential decaying part in its tail, i.e., for $s\gg s_c$. 
$s_c$ decreases with increasing $K$ after $K$ passes $K_c$.
At $K_{p1}$, $\langle S\rangle$ and $s_c$ are equal. 
After $K$ passing $K_{p1}$, 
the mean cluster size is dominated by $s_c$, which makes 
$\bar{s}$ decrease for $K>K_{p1}$. 
However, the mean cluster size increases again as soon as 
$K$ becomes much larger than $K_{p1}$ or $K_c(N)$ because 
the prefactor $K^{1/\mu}$ of the cluster size distribution 
$P(s)$ for $1\ll s\ll s_c$ increases with increasing $K$. 
The mean cluster size decreases only after   
the second peak at $K_{p2}=\mathcal{O}(1)$, where 
$s_c=\mathcal{O}(1)$, as shown in Fig.~\ref{fig:x24} 
as well as  in the exact solution in Fig.~\ref{fig:num_sol}. 
Since $K_{p2}=\mathcal{O}(1)$, a giant cluster 
exists around $K_{p2}$.

\section{Summary and Discussion} 
\label{sec:summary}

In this paper, we have studied the percolation transition of 
the SF random graphs constructed by attaching edges with 
probability proportional to the products of two vertex weights. 
 By utilizing the Potts model representation, 
the giant cluster size, the mean 
cluster size, and the numbers of loops and clusters 
are obtained from the Potts model free energy in the thermodynamic
limit. 
Our general formula for the giant cluster size and the mean 
cluster size are equivalent to those results obtained for a 
given degree sequence if the latter expressions are 
averaged over the grandcanonical ensemble. 
The Potts model formulation allows one to derive other quantities 
such as the number of loops easily. Using this approach, we then 
investigated the critical behaviors of the SF network   
realized by the static model in detail.
 Furthermore, to derive the finite size scaling properties of 
the phase transition, the cluster 
size distribution and the largest cluster size in finite size systems 
are also obtained and used. 
We found that there is a percolation transition for $\lambda=1+1/\mu>3$ 
so that a giant cluster appears abruptly when $K=\langle L \rangle /N$ 
is equal to $K_c$ given by Eq.~(\ref{kc})  
while such a giant cluster is generated gradually without a transition 
for $2<\lambda<3$. Thus the process of formation of the giant cluster
 for the case of  $2<\lambda<3$ is fundamentally 
different from that of $\lambda>3$.

\begin{figure}
\centering
\includegraphics[width=8.0cm]{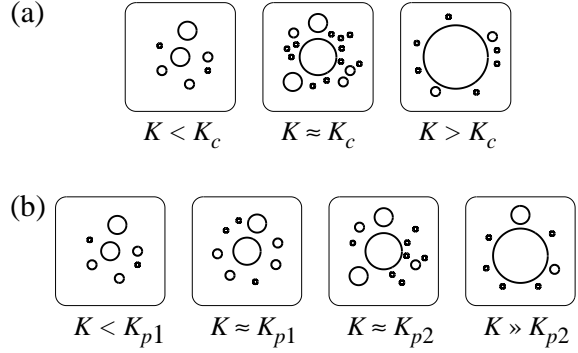}
\caption{Schematic picture for the comparison of cluster evolution
between the case of $\lambda > 3$ (a) and of $2 < \lambda < 3$ (b).}
\label{fig:evolution}
\end{figure}

\begin{figure}
\centering
\includegraphics[width=8.0cm]{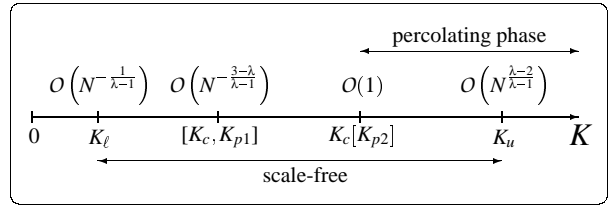}
\caption{Schematic phase diagram of the static model. The SF behavior 
of the degree distribution appears between $K_{\ell}$ and 
$K_u$. 
A giant cluster emerges at $K_c$ for (I,II), and around 
$K_{p2}$ for (III). The quantities in $[\cdots]$ are only for 
(III).}
\label{fig:summary}
\end{figure}

In particular, the difference between the SF graphs with $\lambda>3$ 
and $2<\lambda<3$ is manifested in the mean cluster size. 
For $\lambda>3$, as $K$ or the number of edges increases, many small
clusters grow by attaching edges, which continues up to $K=K_c$, 
and then a giant 
cluster forms by the abrupt coalescence of those small clusters 
as shown in Fig.~\ref{fig:evolution}.
Since we do not count the giant cluster in calculating $\langle s
\rangle$, $\bar{s}$ decreases rapidly as $K$ passes $K_c$. Thus the
mean cluster size exhibits a peak at $K=K_c$, which diverges in
the thermodynamic limit $N\to\infty$.  
On the other hand, for $2<\lambda<3$, the role of $K_c$ is replaced
by $K_{p1}$ which is $ \sim {\mathcal O}
(N^{-(3-\lambda)/(\lambda-1)})$, but after a small peak, 
the mean cluster size increases again after passing $K_{p1}$ as 
seen in Fig.~\ref{fig:x24}: Edges newly introduced either create 
new clusters of size larger than $1$ or merge small clusters to the larger 
one with its size not as large as $\mathcal{O}(N)$. 
Only when $K$ reaches $K_{p2}$, the network is dense enough 
for the giant cluster to swallow up other clusters to reduce 
$\bar{s}$ on adding more edges. The evolution of the static model 
as the density of edges increases is summarized in Fig.~\ref{fig:summary}. 

Due to the characteristics of the power-law degree distribution 
of scale-free graphs, the cluster size distributions exhibit crossover 
behaviors such that they follow different power laws depending on the 
cluster size for given $\lambda$. In particular, for $3<\lambda<4$, 
the scaling exponents $\beta$ and $\bar{\nu}$ depend on the 
degree exponent $\lambda$ continuously while they are equal to 
the conventional mean-field values for $\lambda>4$. 
We have not considered the marginal cases where $1/\mu$ thus 
$\lambda$ is an integer value where logarithmic corrections appear 
as seen in Eq.~(\ref{sigma1int}). 

The static model considered in this paper, though algebraically tractable,
does not have correlations between edges that are important in real
world networks. Extension of this work to the cases allowing correlations
is left for the future work. 

\ack
This work is supported by the KOSEF Grant 
No. R14-2002-059-010000-0 in the ABRL program.

\appendix
\renewcommand{\thesection}{Appendix\ \Alph{section}}
\section{The evaluation of $\Sigma_1(y)$}
\renewcommand{\thesection}{\Alph{section}}
We derive the sums 
\begin{equation}
\Sigma_1(y)\equiv{1\over N}\sum_{i=1}^N e^{-NP_i y},
\end{equation}
where 
$P_i=i^{-\mu}/\zeta_N(\mu)$ $(0<\mu<1)$ with 
$\zeta_N(\mu)=\sum_{i=1}^N i^{-\mu}$ as in Eq.~(\ref{Pi}).

Let us introduce $y_N\equiv {N^{1-\mu}\over \zeta_N(\mu)} y$, 
which is equal to $(1-\mu)y$ in the thermodynamic limit $N\to \infty$ we are 
interested in.          
Then $\Sigma_1(y)=\sum_{i=1}^N s(i)$ with $s(x)=e^{-y_N (N/x)^\mu}/N$. 

First consider the regime where  $y_N N^{\mu}\ll 1$. Then $\Sigma_1(y)$ is 
simply expanded as 
\begin{eqnarray}
\Sigma_1(y)&=&{1\over N}\sum_{n=0}^\infty {(-1)^n \over n!} (y_N N^\mu)^n
\zeta_N(\mu n)\nonumber \\ 
&=&\sum_{n=0}^{\lfloor \frac{1}{\mu}\rfloor} {(-1)^n \over n! (1-\mu n)} y_N^n 
        +{1\over N}\sum_{n=\lfloor \frac{1}{\mu}+1 \rfloor}^\infty 
        {(-1)^n \over n!}(y_N N^\mu)^n \zeta(\mu n),
\label{sigma1o}
\end{eqnarray}
with $\lfloor x \rfloor$ being the greatest integer less than 
or equal to $x$.

On the other hand, in the limit $y_N N^\mu\to \infty$, 
the second summation gives rise to a non-analytic term. 
Since $s(x)$ and its derivatives at $x=1$ and $x=N$ 
have the properties $s^{(n)}(1)=\mathcal{O}(N^{\mu n-1} e^{-y_N N^\mu})$ and 
$s^{(n)}(N)=\mathcal{O}(N^{-n-1})$ with $n\geq 0$, 
the Euler-Maclaurin formula~\cite{arfken} enables 
us to evaluate $\Sigma_1(y)$ in the limit $N\to\infty$ and 
$y_N N^\mu\to\infty$ as follows:
\begin{eqnarray}
\Sigma_1(y)&=&\int_1^N dx\, s(x)
=\int_1^N dx\,{1\over N}e^{-y_N \left(\frac{N}{x}\right)^\mu}\nonumber \\ 
&=&{1\over \mu}y_N^{1\over \mu}\int_{y_N}^\infty dz \, z^{-1-\frac{1}{\mu}} 
e^{-z}  \nonumber \\
&=& {1\over \mu} y_N^{1\over \mu}\Gamma\left(-{1\over \mu},y_N\right),
\label{sigma1a}
\end{eqnarray}
where the incomplete Gamma function $\Gamma(s,x)$ is defined as 
\begin{equation}
\Gamma(s,x)\equiv \int_x^\infty dt \, t^{s-1} 
e^{-t}.
\end{equation}
The incomplete Gamma function can be expressed as  
\begin{eqnarray}
\Gamma(s,x)&=&
\int_x^\infty  dt\, t^{s-1}
\left[e^{-t}-\sum_{n=0}^{\lfloor -s \rfloor} {(-t)^n \over n!}
+\sum_{n=0}^{\lfloor -s \rfloor} {(-t)^n \over n!}\right]
\nonumber \\
&=&
\int_0^\infty dt\, t^{s-1}
\left[\sum_{n=\lfloor -s \rfloor+1}^{\infty} {(-t)^n \over n!}\right]
\nonumber \\
&&-
\int_0^x dt \, t^{s-1}
\left[\sum_{n=\lfloor -s \rfloor+1}^{\infty} {(-t)^n \over n!}\right]
\nonumber \\
&&+\int_x^\infty dt \, t^{s-1}
\left[\sum_{n=0}^{\lfloor -s \rfloor} {(-t)^n \over n!}\right]
\nonumber \\
&=& \Gamma(s)-\sum_{n=0}^\infty {(-1)^n \over n!(s+n)} x^{n+s},
\end{eqnarray}
where it is used that $\Gamma(s)=\int_0^\infty dt \, t^{s-1} [e^{-t}-
\sum_{n=0}^{\lfloor -s \rfloor} (-t)^n/n!]$, 
and therefore,  it follows that 
\begin{eqnarray}
\Sigma_1(y)=\sum_{n=0}^\infty {(-1)^n \over n! (1-\mu n)}
y_N^n-\Gamma\left(1-{1\over \mu}\right)y_N^{1\over \mu}.
\label{sigma1b}
\end{eqnarray}

As $\mu\to 1/m$ with $m$ an integer, a logarithmic term is developed as 
\begin{eqnarray}
&\Sigma_1(y)&=\sum_{n\ne m}^\infty {(-1)^n \over n! (1-\mu n)}
y_N^n \nonumber \\
        &+&{(-1)^m \over (m-1)!} y_N^m \left[
-\ln y_N-\gamma_M +1+{1\over 2}+\cdots {1\over m-1}\right] 
\label{sigma1int}
\end{eqnarray}
since 
\begin{eqnarray}
&&\lim_{\frac{1}{\mu}\to m}\left[ 
{(-1)^m \over m! (1-\mu m)}y_N^m  -\Gamma\left(1-{1\over
                \mu}\right)y_N^{1\over \mu}\right]\nonumber \\
&&={(-y_N)^m \over (1-\mu m)\Gamma(m+1)} \left[
1-{1+(\frac{1}{\mu}-m) \ln y_N \over 
        1+{\Gamma'(m)\over \Gamma(m)}(\frac{1}{\mu}-m)}\right]
        \nonumber \\
&&={(-y_N)^m \over (m-1)!}\left[
-\ln y_N-\gamma_M +1+{1\over 2}+\cdots {1\over m-1}\right], 
\end{eqnarray}
where it is used that $\Gamma(z)\Gamma(1-z)=\pi/\sin \pi z$ and 
$\Gamma'(m)/\Gamma(m)=-\gamma_M+\sum_{n=1}^\infty 
        [1/n-1/(n+m-1)]$.

\renewcommand{\thesection}{Appendix\ \Alph{section}}
\section{Walker algorithm}
Suppose we want to choose discrete random numbers $x_1, x_2,\dots, x_N$,
with probabilities $p_1, p_2,\dots, p_N$, respectively, 
where $p_i$'s are arbitrary yet properly normalized as $\sum_{i=1}^{N}p_i=1$.
Walker algorithm enables one to choose $x_i$'s with appropriate frequencies
by picking a single random number $0\le r<1$.
To do this, however, we have to set up a table $\{(q_i,y_i)\}$.
The table is made in the following way.
\begin{enumerate}
\setlength{\itemsep}{-\parsep}
\item Initialize $q_i$'s as $q_i=Np_i$, ($i=1,2,\dots,N$).
\item Divide $x_i$'s into the poor $(q_i<1)$ and the rich $(q_i>1)$.
\item Pick a poor, say $\mathbf{p}$, and a rich, say $\mathbf{r}$.
\item Fill the shortage of $\mathbf{p}$, $1-q_{\mathbf{p}}$, from $\mathbf{r}$.
\item $\mathbf{p}$ records the donator $\mathbf{r}$ as $y_{\mathbf{p}}=\mathbf{r}$.
\item $q_{\mathbf{r}}$ is updated as $q_{\mathbf{r}}\leftarrow q_{\mathbf{r}}-(1-q_{\mathbf{p}})$.
\item If ${\mathbf{r}}$ becomes poor by the donation, {\em i.e.}, $q_{\mathbf{r}}<1$,
$\mathbf{r}$ enters the list of the poor out of that of the rich.
\item Repeat $(3)$--$(7)$ until there are no poor left (hence the name 
Robin Hood method).
\end{enumerate}
The donate-and-fill steps $(3)$--$(7)$ are performed at most $N-1$ times,
thus the table-making procedure takes time of $\mathcal{O}(N)$.
In its original introduction \cite{walker}, Walker proposed the step 
$(3)$ as ``pick the poorest and the richest,'' which involves additional
sorting operation, increasing the computational cost.
The present scheme follows the implementation of Zaman \cite{robinhood}.
The table-making process can be visualized for a simple $N=3$ case 
as in Fig.~\ref{walker-fig}.
\renewcommand{\thefigure}{B.\arabic{figure}}
\begin{figure}[h]
\centering
\includegraphics[width=8.5cm]{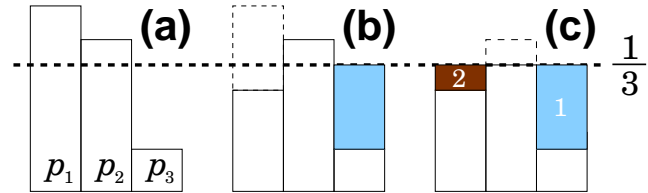}
\caption{Illustration of the table-making procedure in the Walker algorithm
for $N=3$. The heights of the white bars in {\bf (c)} indicates $q_i/N$
and the figures in the shaded boxes $y_i$.
}
\label{walker-fig}
\end{figure}

With the table $\{(q_i,y_i)\}$ at hand, one picks a random number 
$0 \le r <1$. Divide $x=rN+1$ into the integer part $n=\lfloor x\rfloor$ 
and the remainder $d=x-n$. If $d<q_n$, we choose $x_n$, 
otherwise choose $y_n$.
With this scheme one can draw discrete random numbers $x_i$'s 
with appropriate probabilities $p_i$'s.


\begin{thebibliography}{99}
\bibitem{er61}
P.~Erd\H{o}s and A.~R\'{e}nyi, Publ.~Math.~Inst.~Hung.~Acad.~Sci. {\bf 5},
    17 (1960); Bull.~Inst.~Int.~Stat. {\bf 38}, 343 (1961).
\bibitem{ba99}
A.-L.~Barab\'{a}si and R.~Albert, Science {\bf 286}, 509 (1999).
\bibitem{albert02}
R.~Albert and A.-L.~Barab\'{a}si, Rev.~Mod.~Phys. {\bf 74}, 47 (2002).
\bibitem{mendes02}
S. N.~Dorogovtsev and J. F. F.~Mendes, Adv.~Phys. {\bf 51}, 1079 (2002).
\bibitem{newman03}
M. E. J.~Newman, SIAM Rev. {\bf 45}, 167 (2003).
\bibitem{berg02}
J.~Berg and M.~L\"{a}ssig, Phys.~Rev.~Lett. {\bf 89}, 228701 (2002).
\bibitem{manna03} M. Baiesi and S. S. Manna, Phys. Rev. E {\bf 68,} 
047103 (2003).
\bibitem{burda}
Z.~Burda, J. D.~Correia, and A.~Krzywicki, Phys.~Rev.~E {\bf 64}, 
046118 (2001); 
Z.~Burda and A.~Krzywicki, Phys.~Rev.~E {\bf 67}, 046118 (2003).
\bibitem{doro03} S. N. Dorogovtsev, J. F. F. Mendes, and 
A. N. Samukhin, Nucl. Phys. B {\bf 666,} 396 (2003).
\bibitem{farkas}
I.~Farkas, I.~Derenyi, G.~Palla, and T.~Viscek, 
in Lecture notes in
Physics: {\it Networks: structure, dynamics, and function}, edited by
E.~Ben-Naim, H.~Frauenfelder, and Z.~Toroczkai (Springer, 2004).
\bibitem{molloy}
M.~Molloy and B.~Reed, Random Struct. Algorithms {\bf 6}, 161 (1995); 
Combinatorics, Probab. Comput. {\bf 7}, 295 (1998).
\bibitem{newman01}
M. E. J.~Newman, S. H.~Strogatz, and D. J.~Watts, Phys.~Rev.~E {\bf 64}, 
026118 (2001).
\bibitem{goh01}
K.-I.~Goh, B.~Kahng, and D.~Kim, Phys. Rev. Lett. {\bf 87}, 278701
(2001).
\bibitem{caldarelli02}
G.~Caldarelli, A.~Capocci, P.~De~Los~Rios, and M. A.~Mu\~{n}oz, 
Phys.~Rev.~Lett. {\bf 89}, 258702 (2002). 
\bibitem{soderberg02}
B.~S\"{o}derberg, 
Phys.~Rev.~E {\bf 66}, 066121 (2002).
\bibitem{chung02}
F.~Chung and L.~Lu, Annals Combinatorics {\bf 6}, 125 (2002).
\bibitem{aiello02}
W.~Aiello, F.~Chung, and L.~Lu, Exp.~Math. {\bf 10}, 53 (2001).
\bibitem{wu}
P. W.~Kasteleyn and C. M.~Fortuin, 
J.~Phys.~Soc.~Japan Suppl. {\bf 16}, 11 (1969); 
C. M.~Fortuin and P. W.~Kasteleyn, Physica {\bf 57}, 536 (1972).
Also see F. Y.~Wu, Rev. Mod. Phys. {\bf 54}, 235 (1982).
\bibitem{potts_dorogo} 
S. N. Dorogovtsev, A. V. Goltsev, and J. F. F. Mendes, 
cond-mat/0310693.
\bibitem{cohen00}
R.~Cohen, K.~Erez, D.~ben-Avraham, and S.~Havlin, Phys.~Rev.~Lett. {\bf
85}, 4626 (2000); Phys.~Rev.~Lett. {\bf 86}, 3682 (2001).
\bibitem{callaway00}
D. S.~Callaway, M. E. J.~Newman, S. H.~Strogatz, and D. J.~Watts,
Phys.~Rev.~Lett. {\bf 85}, 5468 (2000).
\bibitem{cohen02} R. Cohen, D. ben-Avraham, and S. Havlin, 
Phys. Rev. E {\bf 66}, 036113 (2002).
\bibitem{harris89}
R.~Otter, Ann. Math. Statist. {\bf 20}, 206 (1949);
T. E.~Harris, {\it The Theory of Branching Processes} (Dover, New York, 1989).
\bibitem{aharony92}
D.~Stauffer and A.~Aharony, {\it Introduction to percolation theory}
(Taylor \& Francis, London, 1992).
\bibitem{cohen_book}
R.~Cohen, S.~Havlin, and D.~ben-Avraham, in 
{\it Handbook of Graphs and Networks}, edited by S.~Bornholdt and H.G.~Shuster 
(Willey-VCH, New York, 2002), Chap.~4.
\bibitem{walker} A. J. Walker, Electron. Lett. {\bf 10}, 127 (1974).
\bibitem{robinhood} A. Zaman, unpublished manuscript (1996).
\bibitem{arfken}
G.~Arfken, {\it Mathematical methods for physicists} (Academic, Orlando, 1985).
\end{thebibliography}
\end{document}